\documentclass{jaa}
\usepackage{natbib}
\usepackage{xcolor,hyperref}
\usepackage{url} 
\usepackage{subcaption}
\usepackage{caption}
\usepackage{amsmath}
\usepackage{graphicx}

\begin{document}\sloppy

%%paper title
%%For line breaks \\ can be used within title
\title{Deciphering the spectral properties of the atypical radio relic in A115 using uGMRT, VLA, and LOFAR}

%%author names are separated by comma (,)
%%use \and before the last author name
%%use a * along with the number separated by comma
%% for the  author for correspondence
%%\textsuperscript{number} is used for affiliation
%%\affilOne, \affilTwo etc., upto \affilTwentyfive is possible
%%Please note the first letter after \affil is capitalised in the command
%%

\author{Swarna Chatterjee\textsuperscript{1}, Abhirup Datta\textsuperscript{1}% and AUTHOR3\textsuperscript{2,*}
}
\affilOne{\textsuperscript{1}Department of Astronomy, Astrophysics, and Space Engineering, Indian Institute of Technology Indore, Madhya Pradesh 453552, India}
%\affilTwo{\textsuperscript{2}Department of Q, University Z, Place Pincode, Country.}

%%escape two column mode for title, affiliation and abstract
%%by giving \twocolumn command as shown

\twocolumn[{

\maketitle

%%include \corres to print the corresponding author Email id
\corres{swarna.chatterjee16@gmail.com}

%%include \msinfo for
%%manuscript information such as
%%received, revised and accepted dates
%%
\msinfo{25 April 2024}{11 September 2024}

%%abstract
\begin{abstract}
The mega-parsec scale radio relics at the galaxy cluster periphery are intriguing structures. While textbook examples of relics posit arc-like elongated structures at the clusters' peripheries, several relics display more complex structures deviating from the conventional type. Abell 115 is a galaxy cluster, hosting an atypical radio relic at its northern periphery. Despite the multi-wavelength study of the cluster over the last decades, the origin of the radio relic is still unclear. In this paper, we present a multi-frequency radio study of the cluster to infer the possible mechanism behind the formation of the radio relic. We used new 400 MHz observations with the uGMRT, along with archival VLA 1.5 GHz observations and archival LOFAR 144 MHz observations. Our analysis supports the previous theory on the relic's origin from the passage of a shock front due to an off-axis merger, where the old population of particles from the radio galaxies at the relic location has been re-energised to illuminate the 2 Mpc radio relic.
\end{abstract}

%%insert keywords separated by 3 hyphens using \keywords{words}

}]
%%close the twocolumn escape here

%%include \doinum{number}for the DOI number in the header
%%include \volnum{number} for the volume number in the header
%%include \year{yyyy} for  year of publication in the header
%%include \pgrange{num--num} page range of article in the header
%%include \artcitid{num} for the article citation id
%%include \lp to print last page of the article
%%include \setcounter{page}{pagenum} for the exact starting page of the article

\doinum{12.3456/s78910-011-012-3}
\artcitid{\#\#\#\#}
\volnum{000}
\year{0000}
\pgrange{1--}
\setcounter{page}{1}
\lp{1}

\section{Introduction}
In the hierarchy of structure formation, galaxy clusters form by accumulation of smaller clusters or groups of galaxies. The process of cluster mergers generates a significant amount of energy (10$^{64}$ ergs), driving shock waves and turbulence in the intra-cluster medium (ICM) \citep{Sarazin_2002}. These shocks and turbulence in the disturbed ICM are sites for the acceleration of particles to relativistic speeds and the amplification of magnetic fields within the merging clusters ( see \citealt{Brunetti_2014IJMPD..2330007B, vanweeren_2019SSRv..215...16V, paul_2023JApA} for review). The relativistic charged particles under the influence of the cluster magnetic field emit synchrotron radiation, producing large-scale structures observable in radio frequencies. In the outskirts of the merging clusters, elongated radio structures known as radio relics are observed. These relics often serve as detectable signatures of the weak shock waves ($\mathcal{M}\leq 4$) generated during the merger process \citep{Ensslin_1998A&A}. Despite significant progress in galaxy cluster study at radio wavelength, the exact origins of these radio relics remain inconclusive. While the diffusive shock acceleration (DSA) of thermal particles is considered a plausible explanation for a few relics \citep{Bourdin_2013ApJ, Botteon_2016MNRASB,  chatterjee_2022AJ}, a more commonly accepted scenario involves the re-acceleration of previously accelerated particles within the cluster environment (e.g. \citealt{Gennaro_2018, Botteon_2020AA}). To develop a robust theory on the origins of radio relics, it is essential to gather more samples with detailed information on their origins. This will enhance the current statistical dataset and help determine which mechanisms are most suitable for explaining the origin of radio relics.\\%Studying these intricate sources in greater detail is essential for a more profound understanding of the origin of the relics.
\begin{table*}
	\centering
	\caption{The observation summary for uGMRT and VLA are listed below}
	\label{tab:Table1}
	\hspace{0.01in}
    %\resizebox{\textwidth}{!}
    {
	\begin{tabular}{ccccc} 
	    \hline
	    \hline
		Telescope & Project Code & Frequency & Bandwidth & Time on Source \\
		Configuration & & ( MHz) & ( MHz) & (min) \\
	    \hline
		uGMRT & 35\_095 & 400 & 200 & 78 \\
        uGMRT & 35\_095 & 1260 & 200 & 88 \\
		VLA B & 15A-270 & 1520 & 64 & 40\\
		VLA C & 18B-237 & 1520 & 64 & 137 \\
		VLA D & AF349 & 1520 & 50 & 145 \\
        \hline
        %LOFAR & LOTSS-DR2 & 144 & 43.75 & 15 & $51.32\arcsec  \times 33.55\arcsec $ & $+49.98$\degree & 244\\
	    \hline
	\end{tabular}
	}
\end{table*}
\textbf{Abell 115 (A115)} is a massive ($M_{\rm SZ} = 7.2 \pm 0.5 \times10^{14}\;M_{\odot}$) 
merging cluster at redshift $z=0.197$ \citep{Planck_Collaboration_2014AA}. The cluster has been widely studied in the last decades due to its distinctive features. Using Einstein imaging observations, \citet{Forman_1981ApJ} identified an asymmetric double-peaked structure in the X-ray surface brightness (SB) map. Additionally, the presence of two sub-clusters roughly coincident with the X-ray peaks was observed with optical and weak lensing analysis of the cluster \citep{Barrena_2007AA, Kim_2019ApJ}. The trailing gas emission from the subclusters and the hot region between the cluster observed from X-ray observations suggest that A115 is in a post-merging binary system. Using VLA 1.4 GHz observations, \citet{Govoni_2001AA} reported an atypical radio relic at the north of the cluster, with the relic being detected at both high (15$^{\prime\prime}$) and low resolution (35$^{\prime\prime}$). The morphology of the relic exhibits a rough division into two parts, accompanied by a distinctive peculiarity. The western emission aligns with the periphery of the northern sub-cluster, while the eastern portion of the relic extends well beyond the X-ray emitting region of the cluster.
%\citep[][; see left panel of Fig.~\ref{fig: A2108_2693_2680_images}].

There are two competing ideas about the origin of this relic. {\citet{Govoni_2001AA} suggested} that the relic originates from particles ejected from the radio galaxies at the north of the cluster. In contrast, \citet{Hallman_2018ApJ}, based on numerical simulations, suggested that during an off-axis merger of the two sub-clusters, a bow shock was created driven by the movement of the cold sub-clusters, which is the origin of the formation of the radio relic. However, numerical simulation with weak lensing masses indicates that even with a rich population of fossil electrons, a shock would be insufficient to produce such a large radio relic \citep{Lee_2020ApJ}. The morphology of the western part of the relic is in line with numerical simulations of an off-axis merger \citep{Hallman_2018ApJ,Lee_2020ApJ}. \citet{Botteon_2016MNRAS} detected a shock co-spatially located to the western part of the relic using \textit{Chandra} X-ray observations. However, due to the lack of X-ray emission in the eastern region and the lack of a detailed spectral index study, there is no strong evidence in support of the origin of the relic in the cluster in such an unusual position. Here, we present a spectral index study of the radio relic in A115 using new uGMRT, along with archival VLA and LOFAR observations.\\
 We have assumed a $\Lambda$CDM cosmology with $\Omega_m = 0.3$, $\Omega_{\Lambda}=0.7$ and $H_0=70$~km s$^{-1}$~Mpc$^{-1}$. At the cluster redshift of $z = 0.192$, $1^{\prime\prime}$ corresponds to 3.19 kpc.

\section{Radio Observations and Data Reduction}

\begin{figure}
\includegraphics[width=0.9\columnwidth]{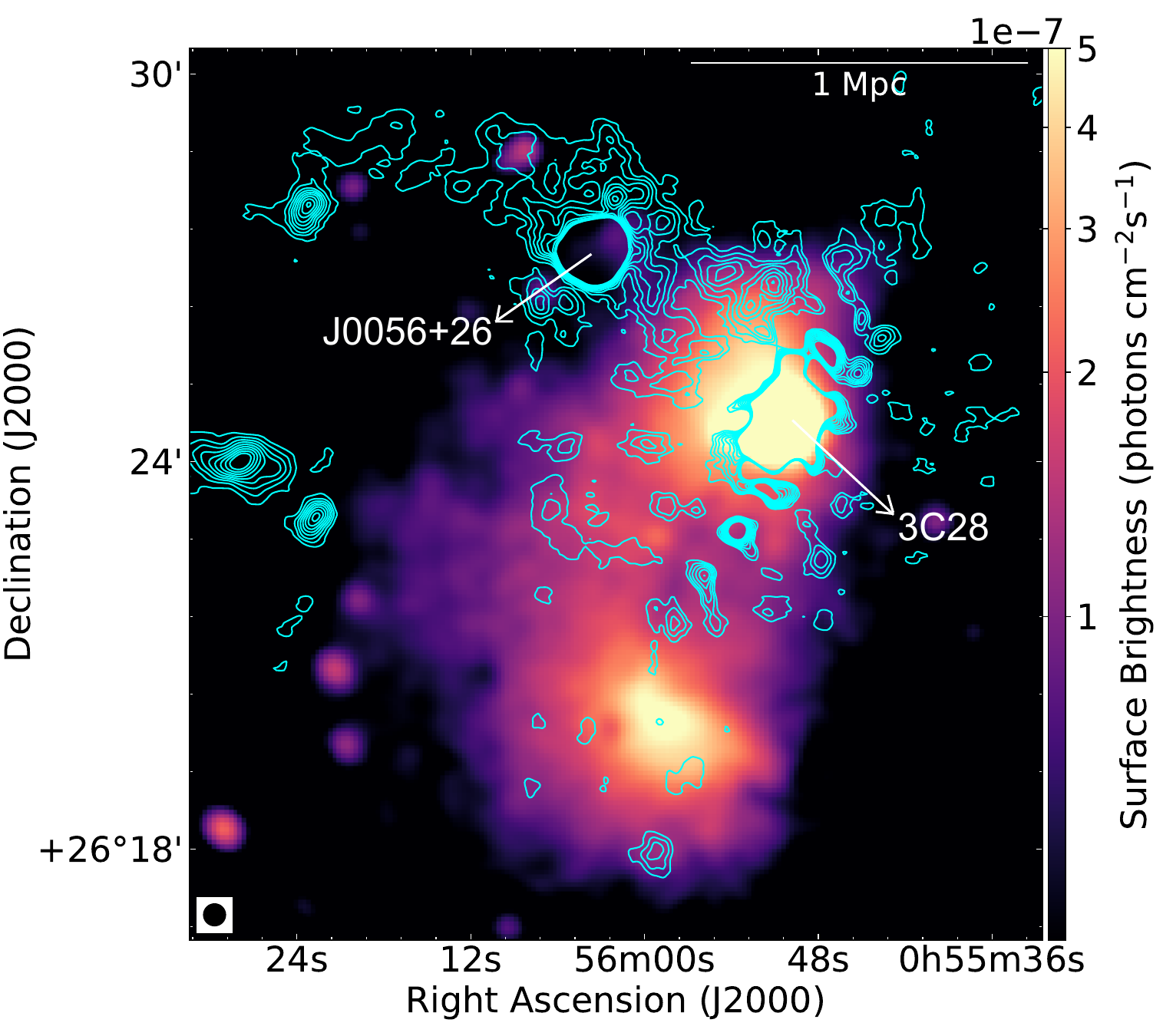} 
\caption{A115 XMM Newton X-ray image overlaid with uGMRT 400 MHz image contours in cyan. The restoring beam of radio image is $20^{\prime\prime}  \times 20^{\prime\prime} $, P.A 0$^\circ$ and the contours are placed at $(3,6,9,...)\times 0.6$ mJy/beam}
\label{fig: xray_radio}
\end{figure}
We used uGMRT 400 MHz and 1.26 GHz observation of the cluster. Moreover, we used archival VLA B, C and D array data at 1.5 GHz and LOFAR 144 MHz images from \citep{Botteon_2022AA} to understand the spectral behaviour of the relic in A115. The observation strategy for uGMRT and VLA are detailed in Table \ref{tab:Table1}.

\subsection{uGMRT Observation}
We observed A115 with the aim of probing its spectral properties with uGMRT band-3 (central frequency 400 MHz) and band-5 (central frequency 1260 MHz). 3C48 was used as the flux calibrator, and 0119+321 as the phase calibrator.

\begin{figure*}
    \includegraphics[width=0.485\textwidth]{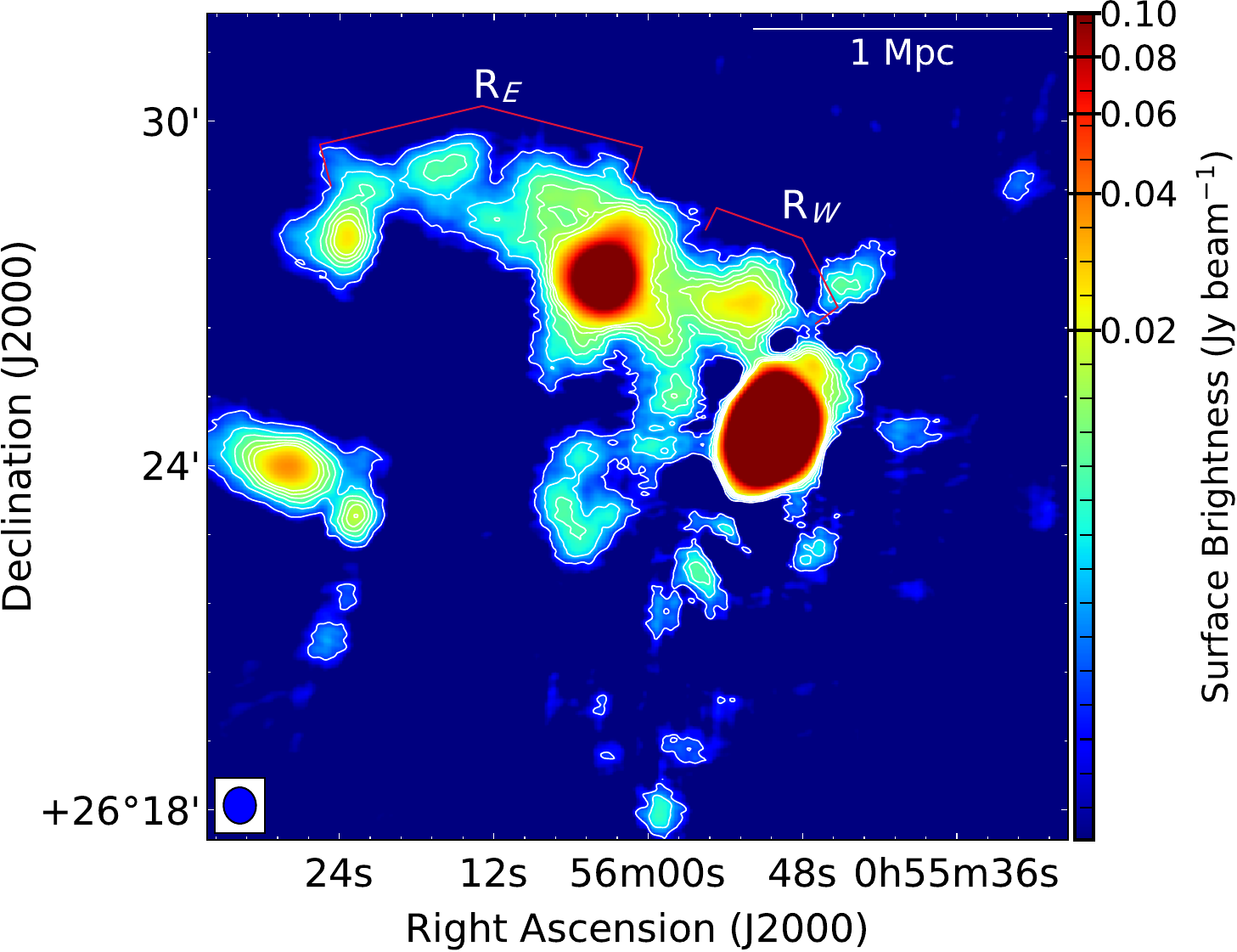}
    \includegraphics[width=0.5\textwidth]{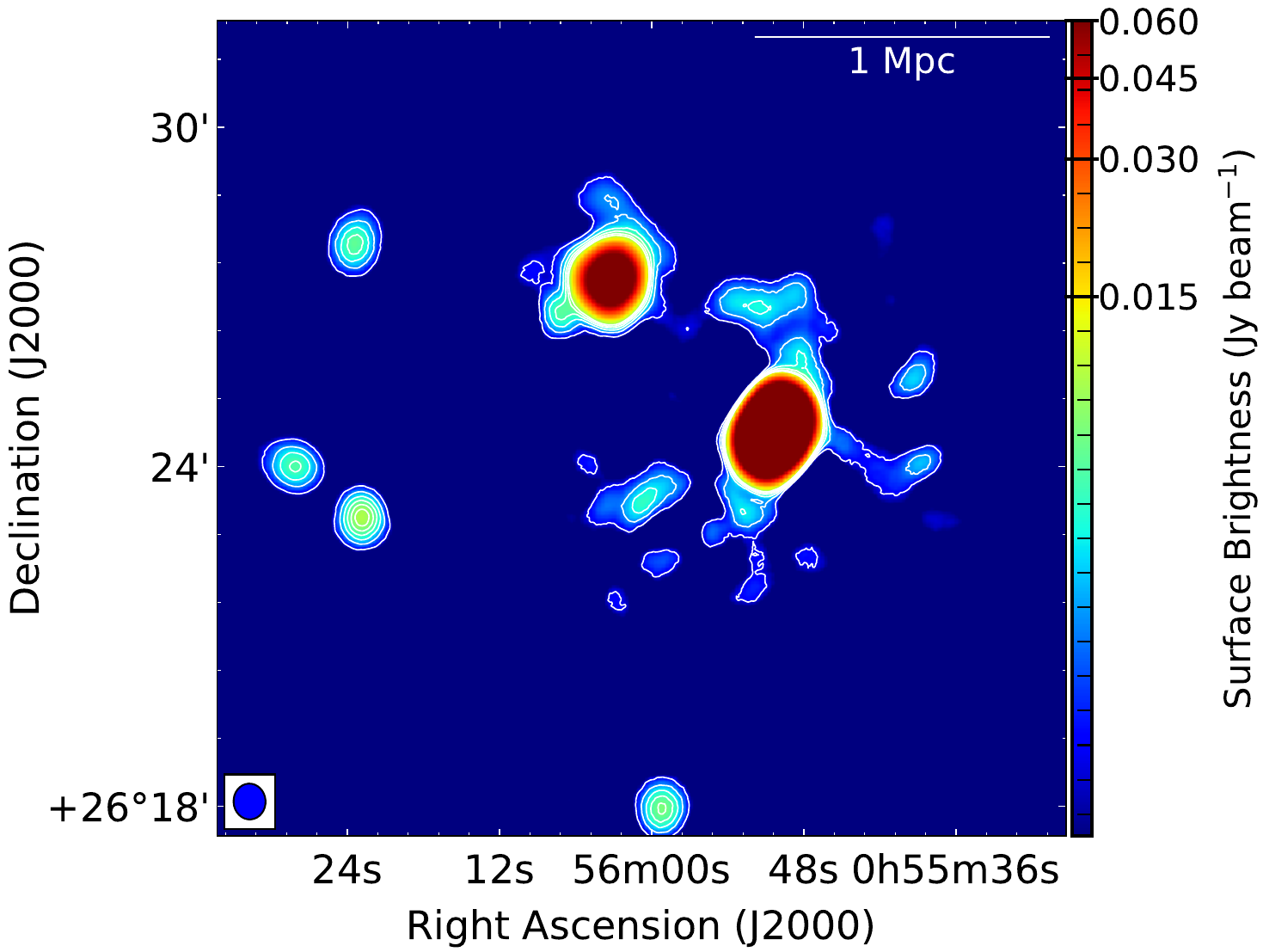}\\
    \includegraphics[width=0.5\textwidth]
    {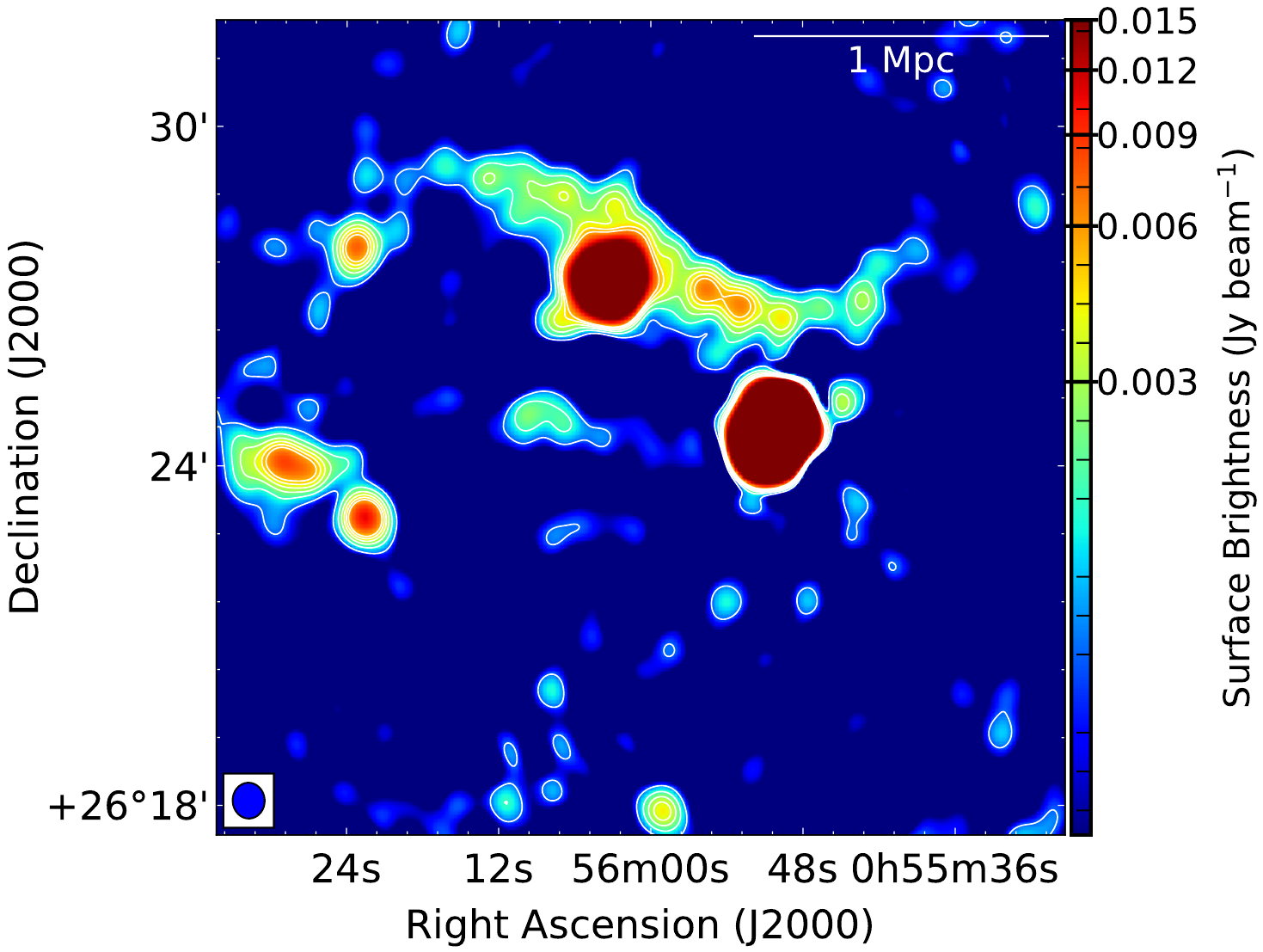}
    \includegraphics[width=0.5\textwidth]{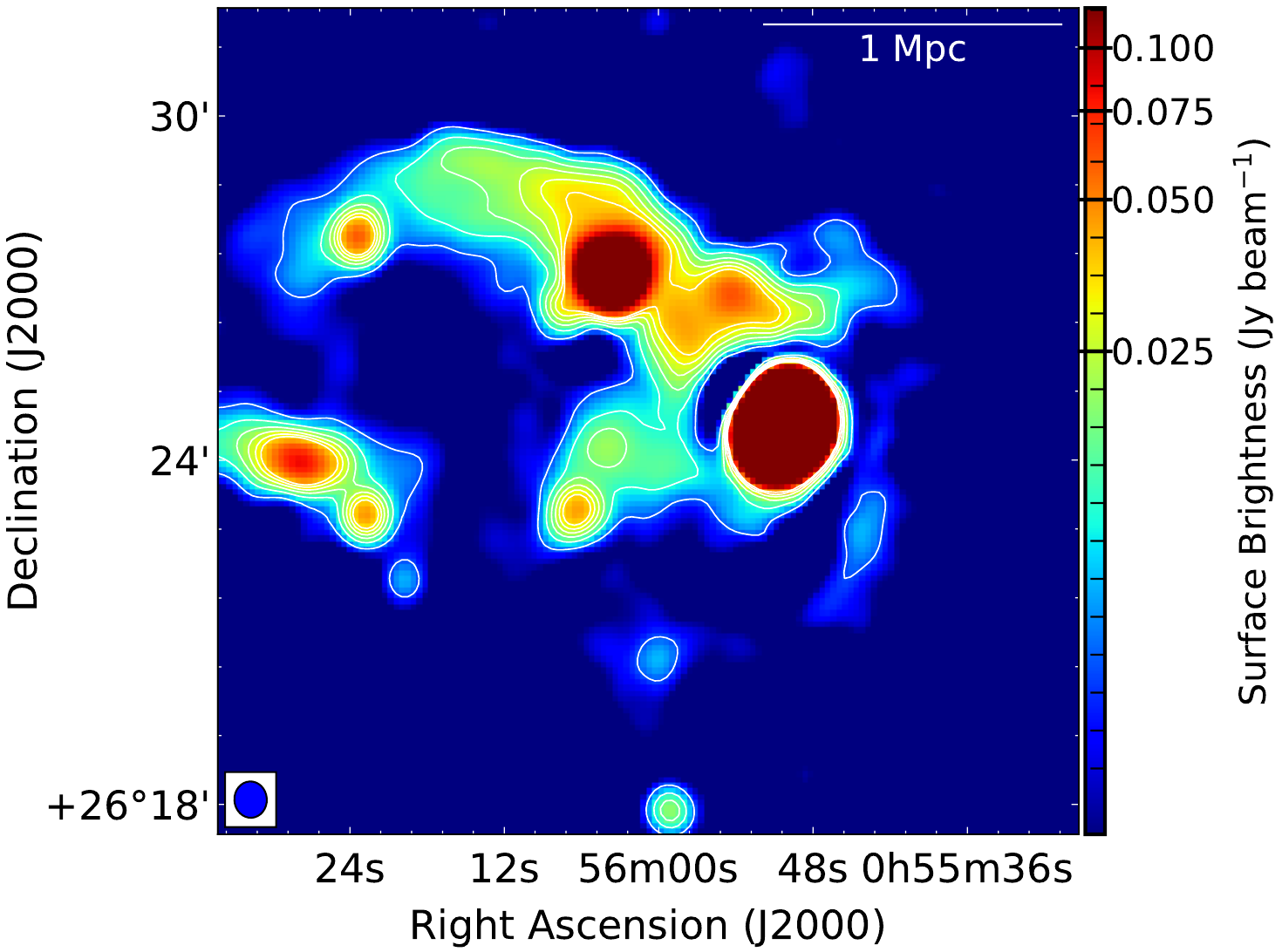}
\caption{A115 images with restoring beam $38.0^{\prime\prime}  \times 33.9^{\prime\prime} $, P.A 3.2$^\circ$; with respective white contours overlaid. Top left: uGMRT 400 MHz image. The contour levels increase as  $(3,6,9,...)\times 1$ mJy/beam. Top right: uGMRT 1.26 GHz image. The contour levels increase as  $(3,6,9,...)\times 0.5$ mJy/beam.  Bottom Left: Combined image from VLA B, C and D array at 1.5 GHz. The contour levels increase as  $(3,6,9,...)\times 0.3$ mJy/beam. Bottom right: LOFAR 144 MHz image. The contour levels increase as  $(3,6,9,...)\times 2$ mJy/beam.}
\label{fig: A115_radio}
\end{figure*} 
The uGMRT data reduction was done using the Source Peeling and Atmosphere Modelling (SPAM) pipeline \citep{Intema_refId0}. The 200 MHz wide bandwidth of band-3 data was split into six subbands in SPAM, and thereafter, the standard procedure, including direction-independent calibration, flagging of bad data and direction-dependant calibration, was performed in the sub-bands. The calibrated outputs were further used for imaging. Despite the direction-dependent approach, the presence of sidelobes from the strong radio source 3C28 affects the emission. To bring out the diffuse emission more, we created the final images using an uvrange of 0.2 - 41k$\lambda$ and using Brigg's robust 1 in WSClean. Briggs robust 1 weighting helps to emphasize short baselines in the image, thus highlighting larger-scale features. %Therefore, we used 15 arcsec gaussian tapering (50kpc at cluster redshift) following \citep{Botteon_2022AA} to bring out the diffuse radio emission. The final images were created with restoring beam size of 30$^{\prime\prime}$ . Fig ~\ref{fig: A115_radio} shows the radio images of the cluster at uGMRT 400 MHz and 1250 MHz.

The band-5 data was also analysed in the same manner in SPAM. Despite multiple trials, the band-5 uGMRT image could not reveal the relic emission much. Therefore, we use the 1.5 GHz VLA image of the cluster.

\subsection{VLA Observation:}
We have used VLA observation of the cluster from B, C and D array. The VLA data was reduced using Common Astronomy Software Application (CASA\footnote{\url{https://casa.nrao.edu/}}), where the different configuration data were treated separately for RFI flagging and direction-independent calibration. The flux density was set using \citet{scaife_10.1111/j.1745-3933.2012.01251.x}. A more detailed data reduction procedure of the B and D array is presented in \citet{Hallman_2018ApJ}. %Additionally, we analysed the C configuration data where the traditional calibration, flagging and self calibration was done using CASA. 
Next, the target observation from each array was combined using the CASA task \textit{concat} to create the image of the cluster. The final images were created using the common uvrange of 0.2 - 41k$\lambda$ in WSClean. \\

\subsection{LOFAR Images:}
We used the 144 MHz LOFAR images of A115 from the LOFAR Two-metre Sky Survey data release 2 (LOTSS DR2)\footnote{\url{https://lofar-surveys.org/planck\_dr2.html}}. We used the low-resolution image of the cluster from the survey, and the same is shown in Figure \ref{fig: A115_radio} (bottom right). A detailed analysis and imaging procedure is given in \citet{Botteon_2022AA}.
\begin{figure}
\includegraphics[width=0.45\textwidth]{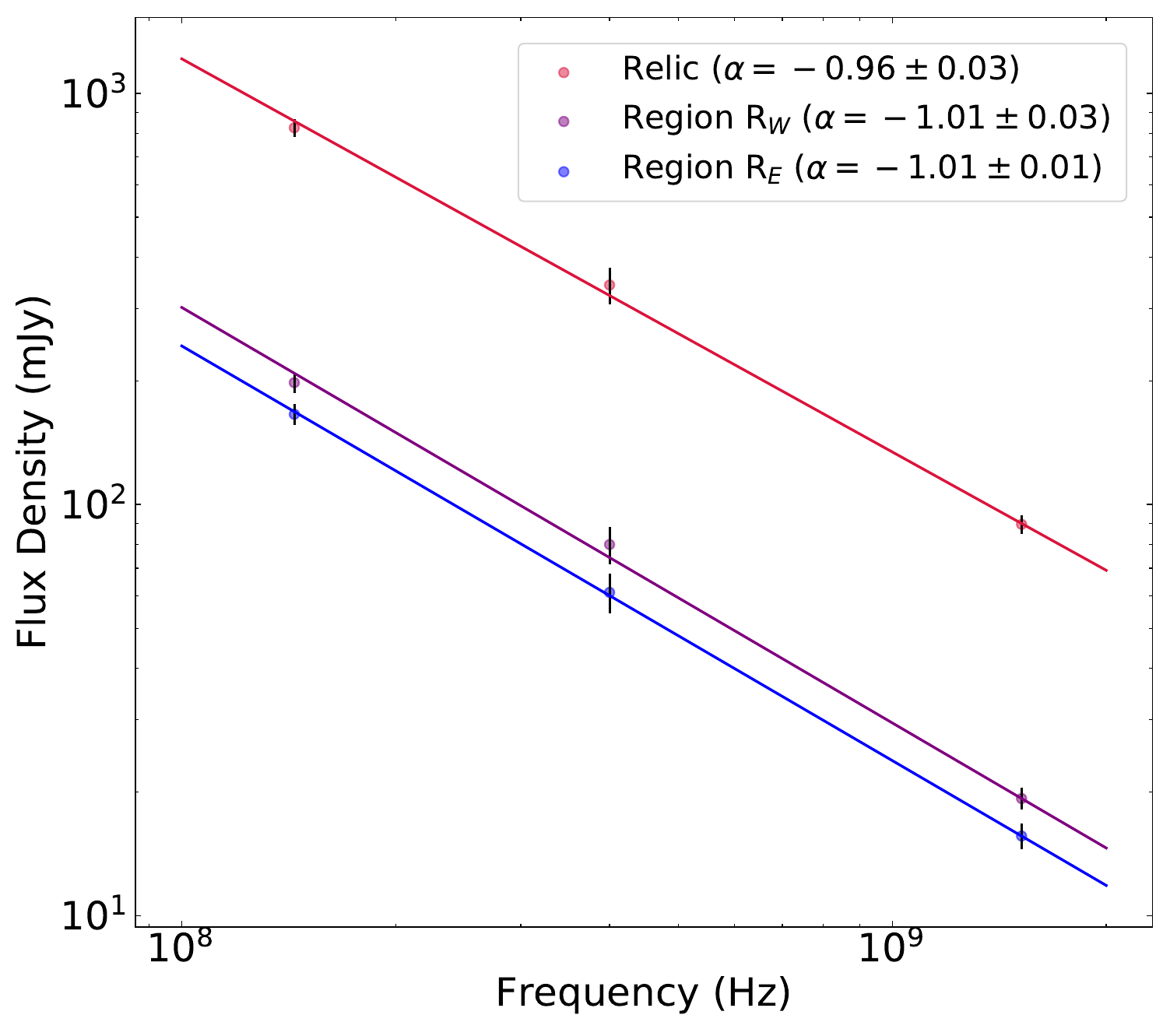}
\caption{Power law fitting for the entire relic emission (R$_W$+R$_E$ emission), the region R$_W$ and region R$_E$.}
\label{fig: powlow}
\end{figure}

\begin{figure*}
\includegraphics[width=0.48\textwidth, height=18em]{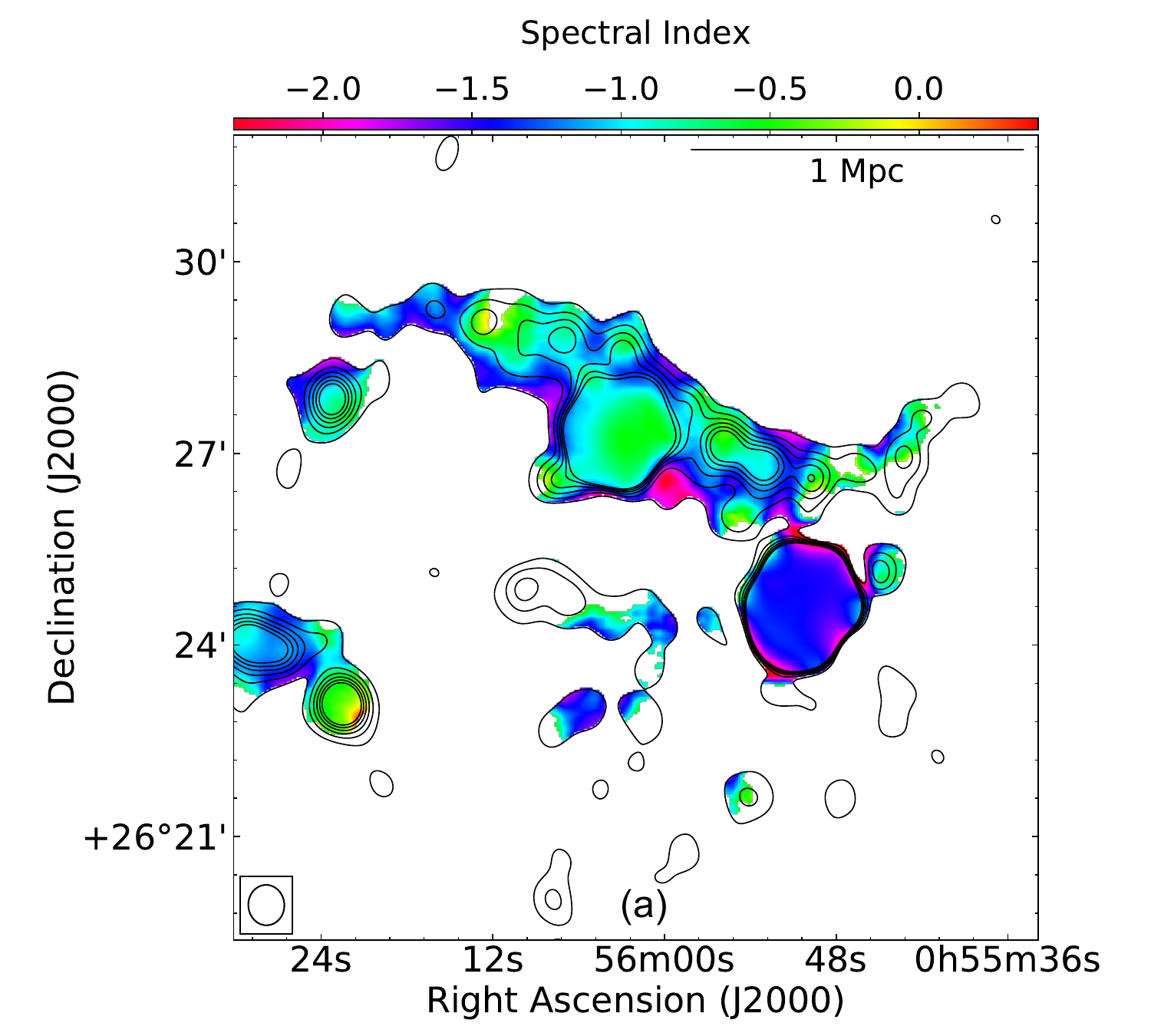}
\includegraphics[width=0.48\textwidth, height=18em]{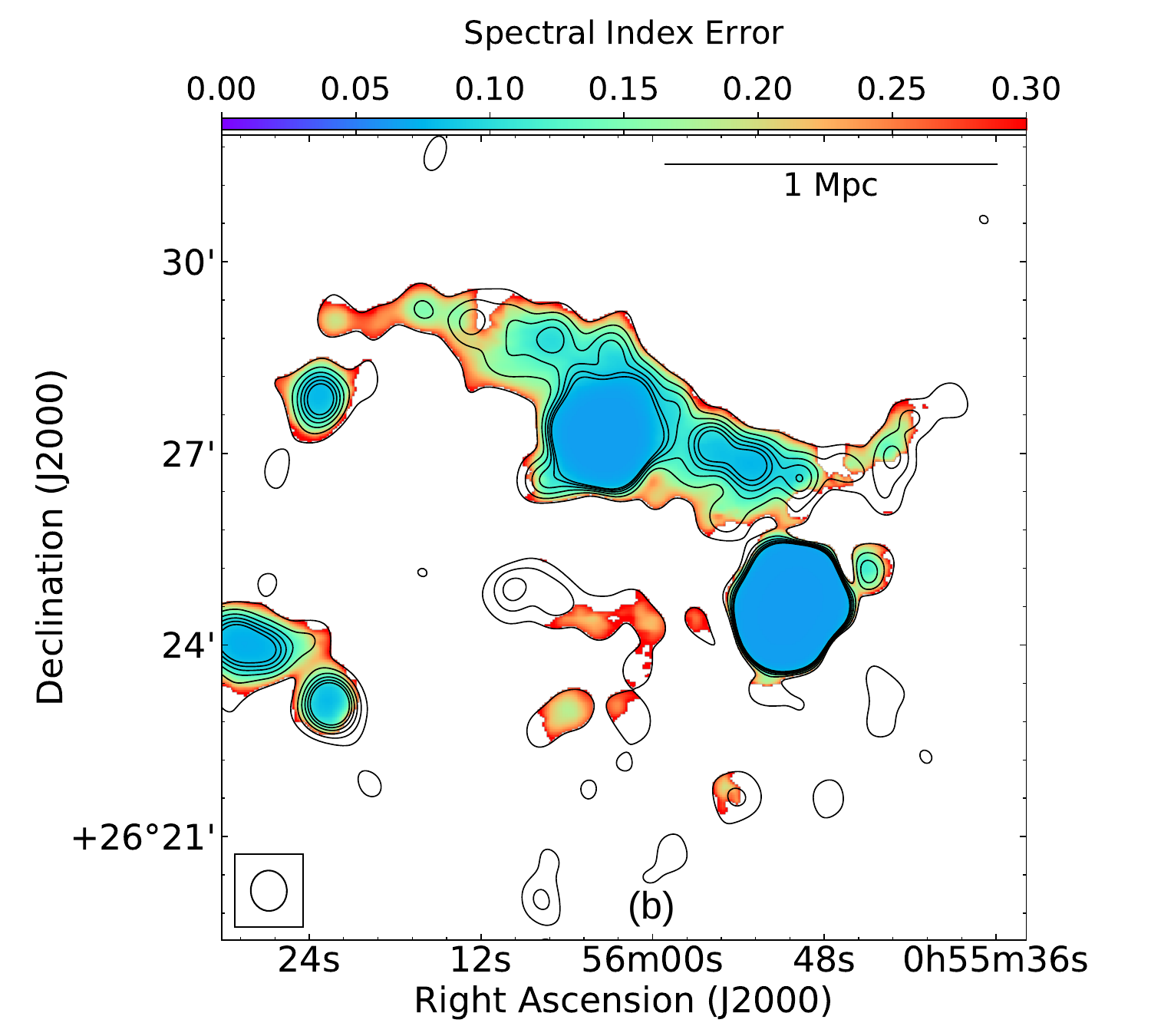}
\includegraphics[width=0.48\textwidth, height=18em]{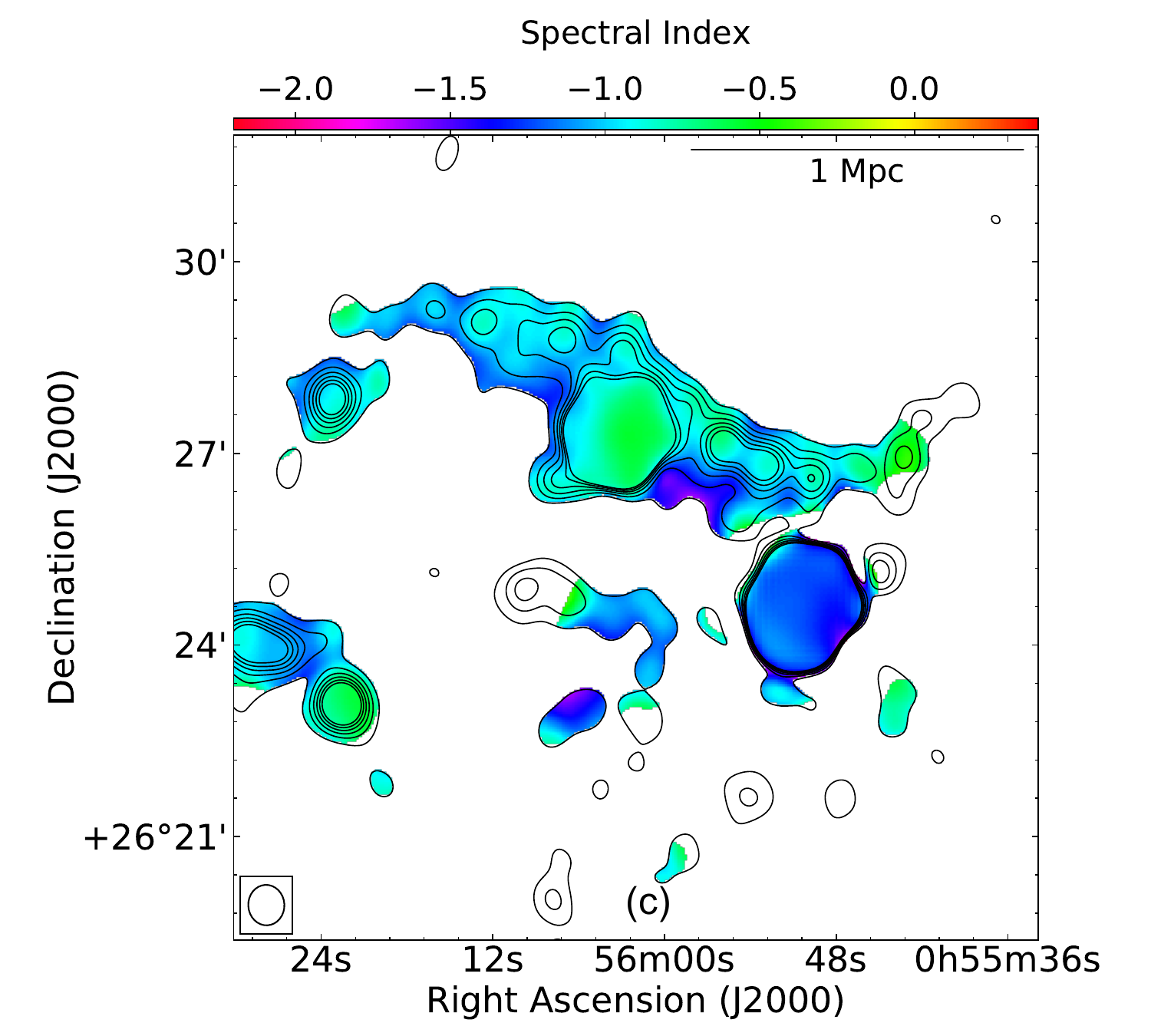}
\includegraphics[width=0.48\textwidth, height=18em]{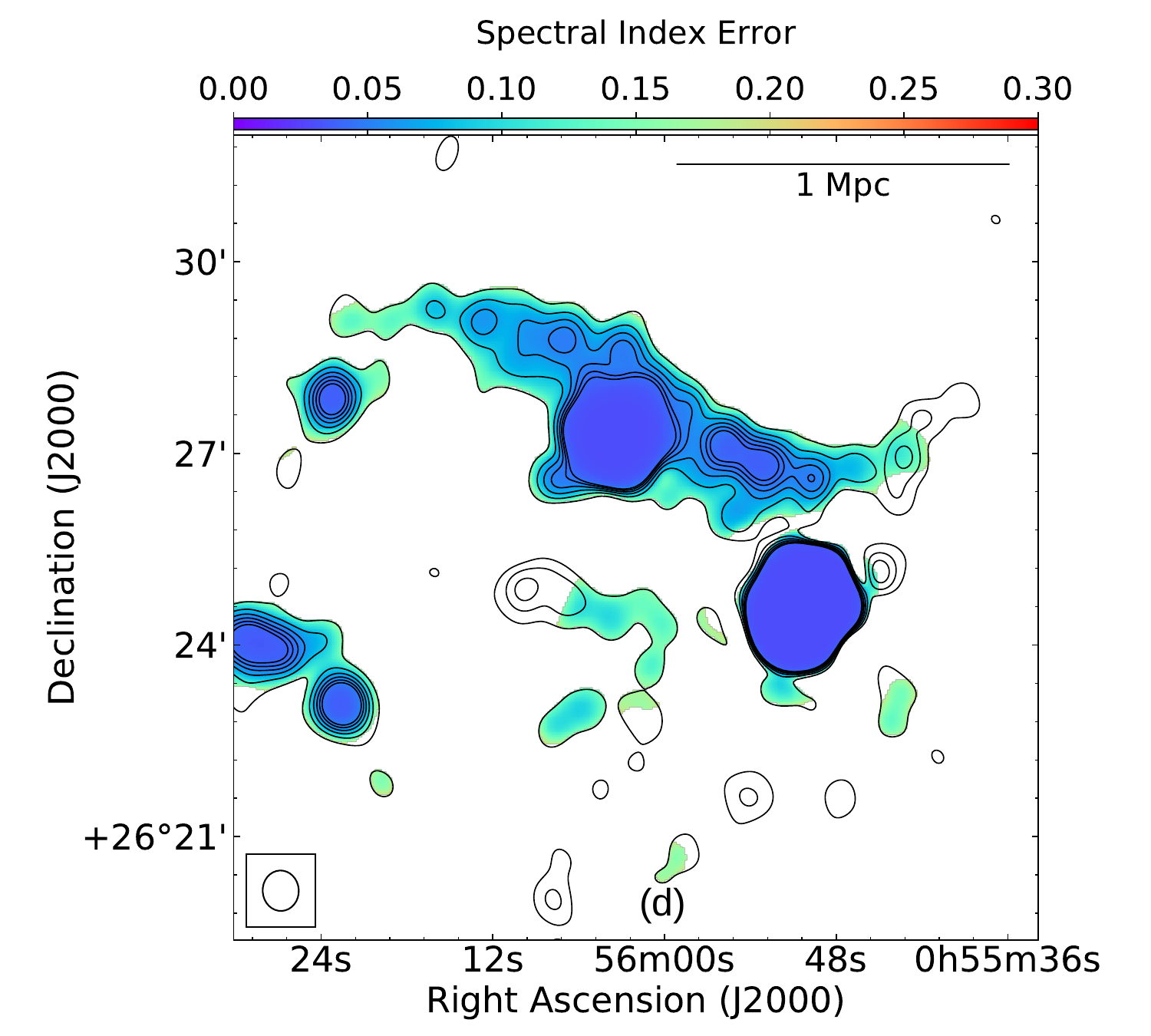}
\includegraphics[width=0.49\textwidth, height=19em]{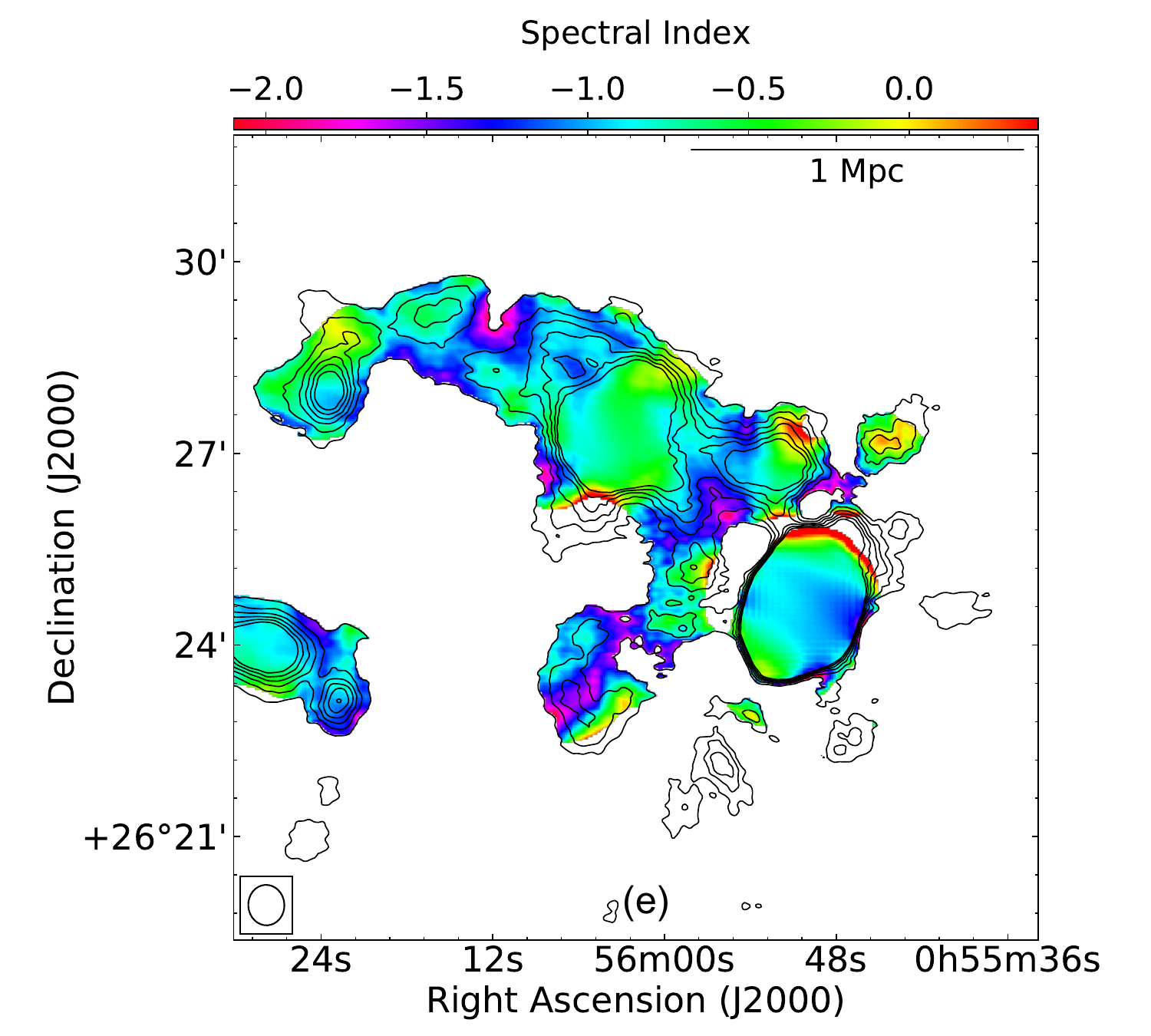}
\includegraphics[width=0.49\textwidth, height=19em]{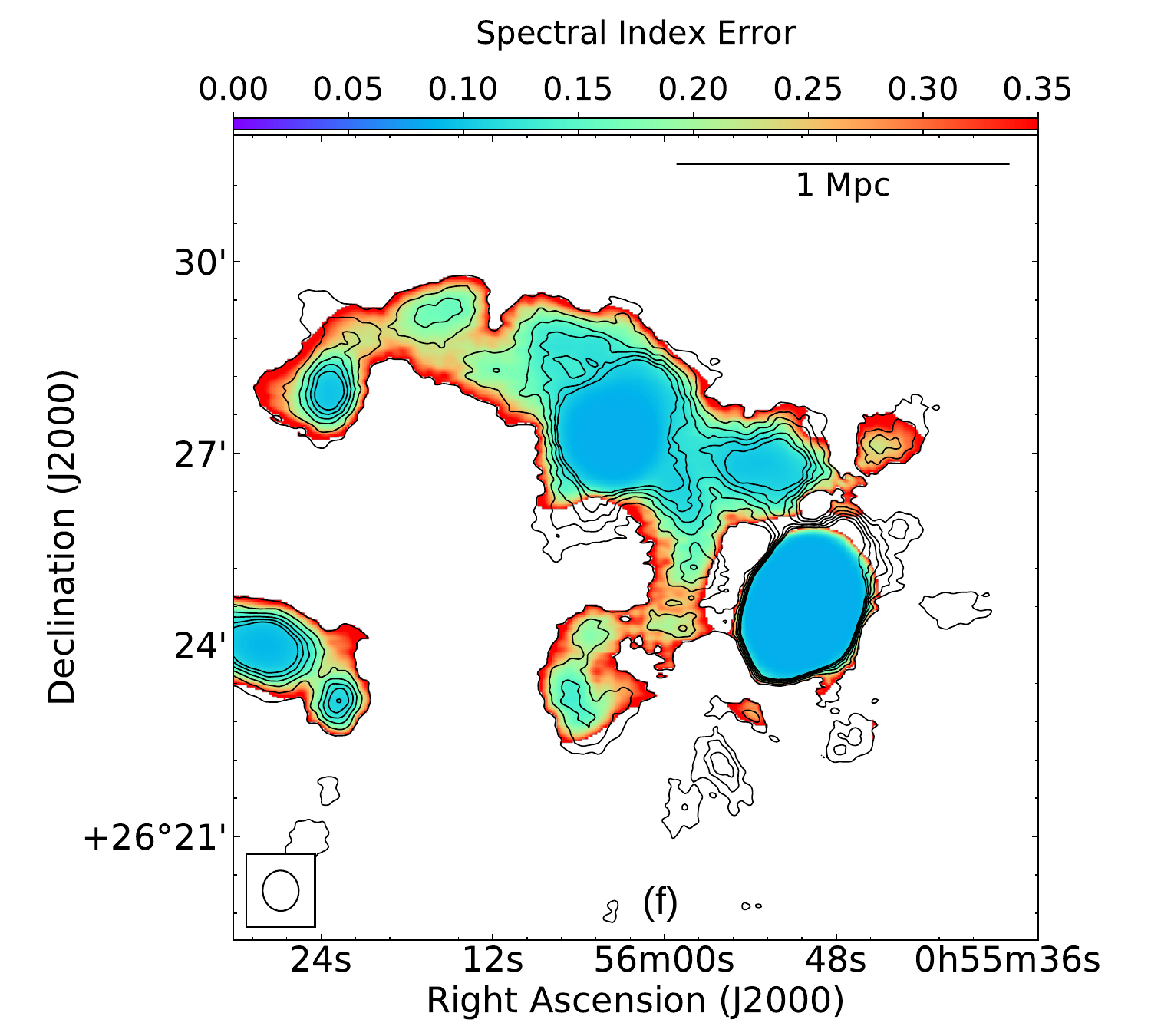}
\caption{ A115 spectral index map with restoring beam $38.0^{\prime\prime}  \times 33.9^{\prime\prime} $, P.A 3.2$^\circ$. Top: (a) Spectral index map and (b) spectral index error map between VLA 1.5 GHz and uGMRT 400 MHz overlaid with 1.5 GHz radio contours in black. Middle: (c) spectral index map and (d) spectral index error map between 1.5 GHz and LOFAR 144 MHz overlaid with 1.5 GHz radio contours in black. Bottom: (e) spectral index map and (f) spectral index error map between 400 MHz and 144 MHz overlaid with 400 MHz radio contours in black.}
\label{fig: spix1}
\end{figure*}

\section{Results}

%Fig. 1 shows the presence of three radio galaxies in the vicinity of the relic in A115. 
Figure~\ref{fig: xray_radio} shows the XMM Newton X-ray map of the cluster overlaid with the uGMRT 400 MHz image contours in cyan created using WSClean v.3.1.0 \citep{Offringa_2017MNRAS} with Brigg's robust 0.5 and convolved with a beam $20^{\prime\prime} \times 20^{\prime\prime} $, P.A 0$^\circ$. The brightest cluster galaxy (BCG) of the northern subcluster, 3C28 and the narrow-angle tail (NAT) galaxy, J0056+26 are marked in the figure. The image fails to capture the relic emission in the western part entirely. Consequently, for further spectral studies, we created low-resolution images convolved with a beam size of $38.0^{\prime\prime} \times 33.9^{\prime\prime}$, P.A. 3.2$^\circ$, which matches the resolution of the LOTSS DR2 low-resolution image of the cluster. Figure~\ref{fig: A115_radio} (Top Left) represents the uGMRT 400 MHz image of the cluster with an image noise, $\sigma = 1.0$ mJy/beam. The relic emission spans over 1.8 Mpc $\times$ 0.45 Mpc at 400 MHz. The western and eastern parts of the relic emission are marked as R$_W$ and R$_E$, respectively. As mentioned earlier, we do not obtain much of the relic emission in the uGMRT band-5 observation (Figure \ref{fig: A115_radio} Top Right).
Figure ~\ref{fig: A115_radio} (Bottom Left) shows the combined VLA image of the cluster. The relic expands over 1.5$\times$ 0.3 Mpc at 1.5 GHz. A trailing emission from the radio galaxy J0056+26 is observed near the R$_E$ emission at 1.5 GHz. The same trailing signature is also observed in the LOFAR 144 MHz image. The largest expansion of the relic was observed at 144 MHz, to be 2 Mpc $\times$ 0.5 Mpc (Figure ~\ref{fig: A115_radio} Bottom Right). The western part of the relic shows a very bright emission in all the images. However, the emission in the eastern part is comparatively faint, and is not observed at 1.26 GHz.  
%Due to the presence of the sidelobes from the strong radio source 3C28 at the western part of the relic emission, the emission is heavily affected in the uGMRT 400MHz observation. %The eastern part of the relic is more expanded at 400 MHz. 
\subsection{Flux density and spectral index estimation}
The presence of the NAT galaxy in the midst of the relic makes the compact source subtraction difficult. Therefore, to measure the integrated flux density of the relic, we measured the flux density of the entire relic regions encompassing emission from R$_E$ and R$_W$ regions and the galaxies. Next, we created high-resolution diffuse emission free images of the cluster with uvrange $> 1.5\mathrm{k}\lambda$ at 400 MHz and 1.5 GHz and calculated the compact source flux densities using PyBDSF. For the LOFAR image, the compact source flux densities were calculated using the high-resolution image available in the LOTSS DR2 image archive. The radio galaxy fluxes were subtracted from the total flux density of the relic region. The measured flux densities of the relic are, S$_{144 \mathrm{MHz}} = 1294.8 \pm 66.9$  mJy, S$_{400 \mathrm{MHz}} = 456.3 \pm 46.1$ mJy 
and S$_{1.5 \mathrm{GHz}} = 103.5 \pm  5.5$ mJy. The uncertainty in flux density estimation was calculated considering a 5\% flux calibration error at 144 MHz \citep{Botteon_2022AA} and 1.5 GHz \citep{Botteon_2016MNRAS} and 7\% flux calibration error at 400 MHz \citep{Raja_2023arXiv,Chatterjee_2024MNRAS}.% and  flux using the equation
%\begin{equation} \label{eq1}
%\Delta S=\sqrt{(\sigma_\mathrm{cal}\ S)^2+(\sigma_\mathrm{rms} \sqrt{N_\mathrm{beam}})^2} 
%end{equation} 

Spectral index measurement requires information on flux densities from the same region of emission. We estimated the integrated spectral index of the relic over the region, showing emission in all three frequencies, using the following equation.
\begin{equation}\label{eq1_s}
    \alpha=log(S_1/S_2)/log(\nu_1/\nu_2)
\end{equation}
 where S$_i$ denotes flux densities at different frequencies and $\nu_i$ denotes respective frequencies. \\
  %Thus, before proceeding further, we inspected for any misalignment between the 400 MHz and 1520 MHz images. A misalignment of 0.21$^{\prime\prime}$ in R.A and 4.5$^{\prime\prime}$ in Dec was found, which was thereafter corrected. 
We found that the integrated spectrum of the relic follows a power law with spectral index $\alpha= -0.96\pm 0.03$ over 144 MHz to 1.5 GHz (Figure~\ref{fig: powlow}). Here, we note that the presence of compact sources in the image can affect the flux density measurements. Thus, we calculated the flux densities of the region R$_W$ and R$_E$ separately, excluding the regions encompassing the radio galaxies. The R$_W$ and R$_E$ region follows a power law with $\alpha= -1.01\pm 0.01$ and $\alpha= -1.01\pm 0.03$, respectively. Previously \citet{Hallman_2018ApJ} estimated an integrated spectral index -1.1$\pm$0.2 for the R$_W$ emission which aligns with our findings for the R$_W$ region. 

\begin{figure*}
\includegraphics[width=\columnwidth]{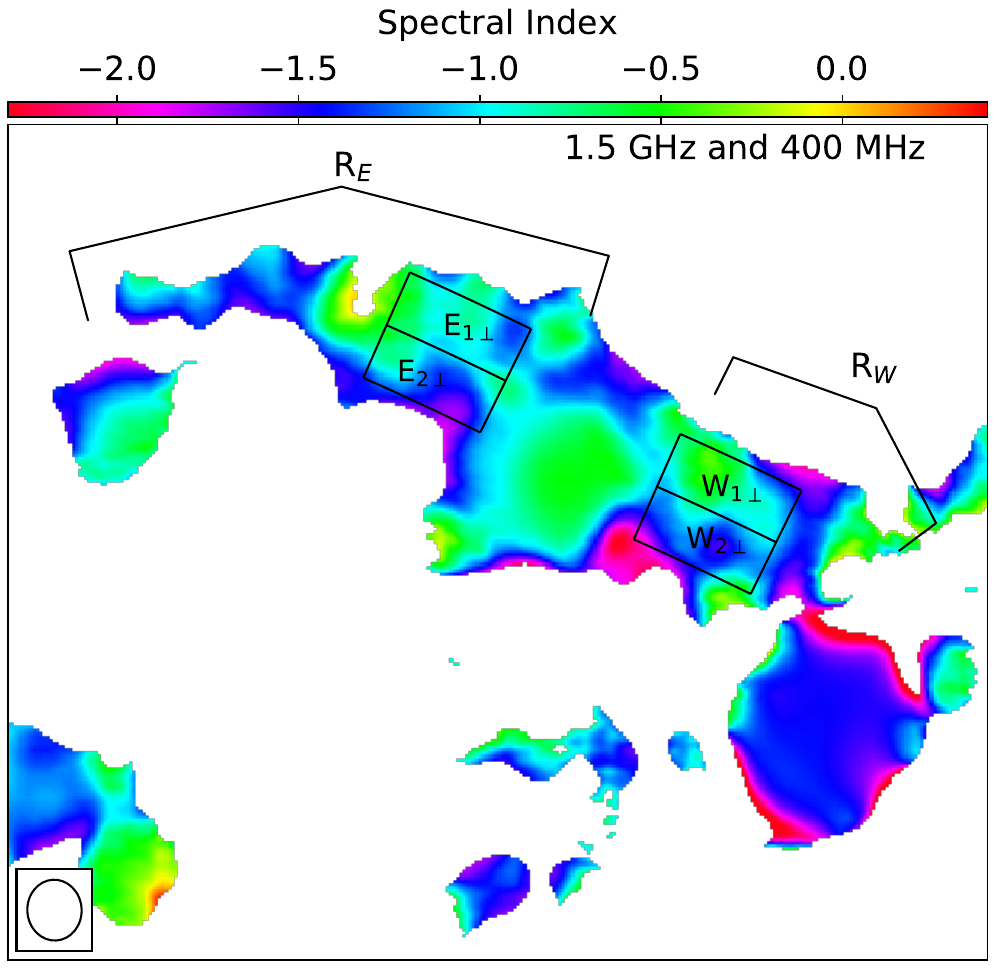}
\includegraphics[width=\columnwidth]{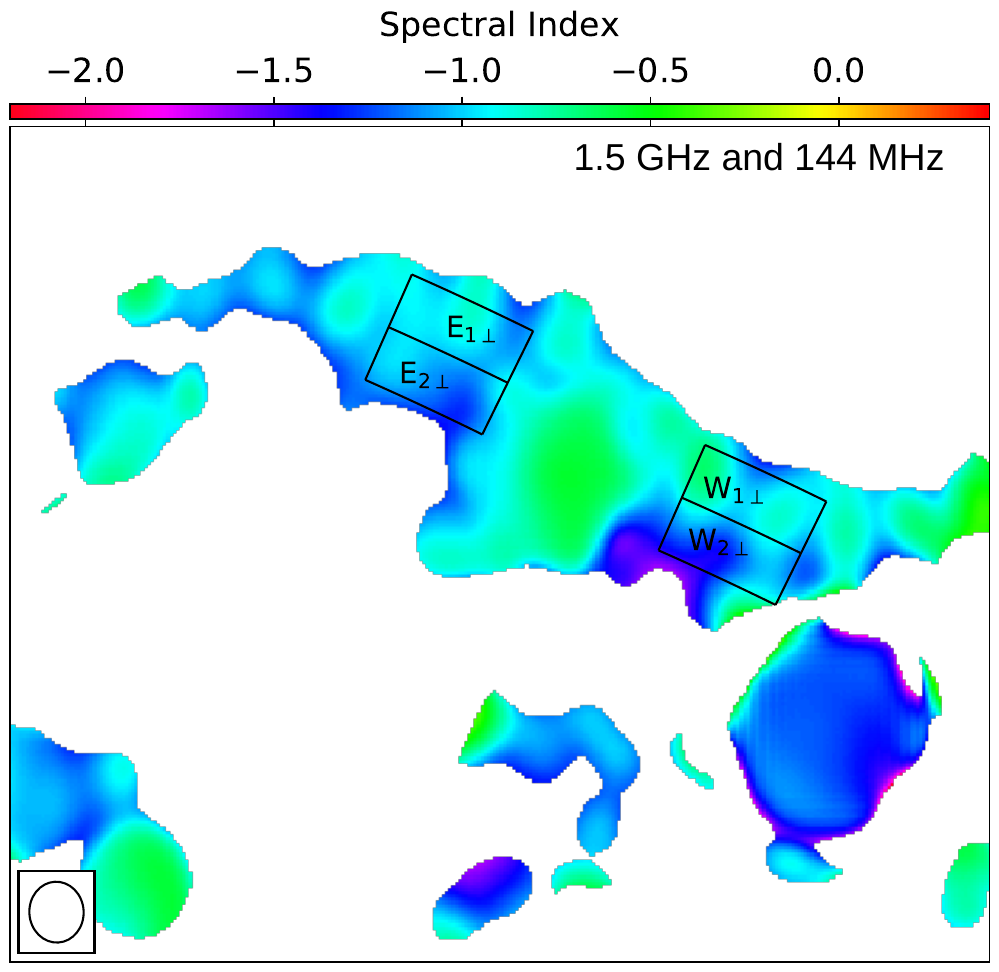}\\
\includegraphics[width=\columnwidth]{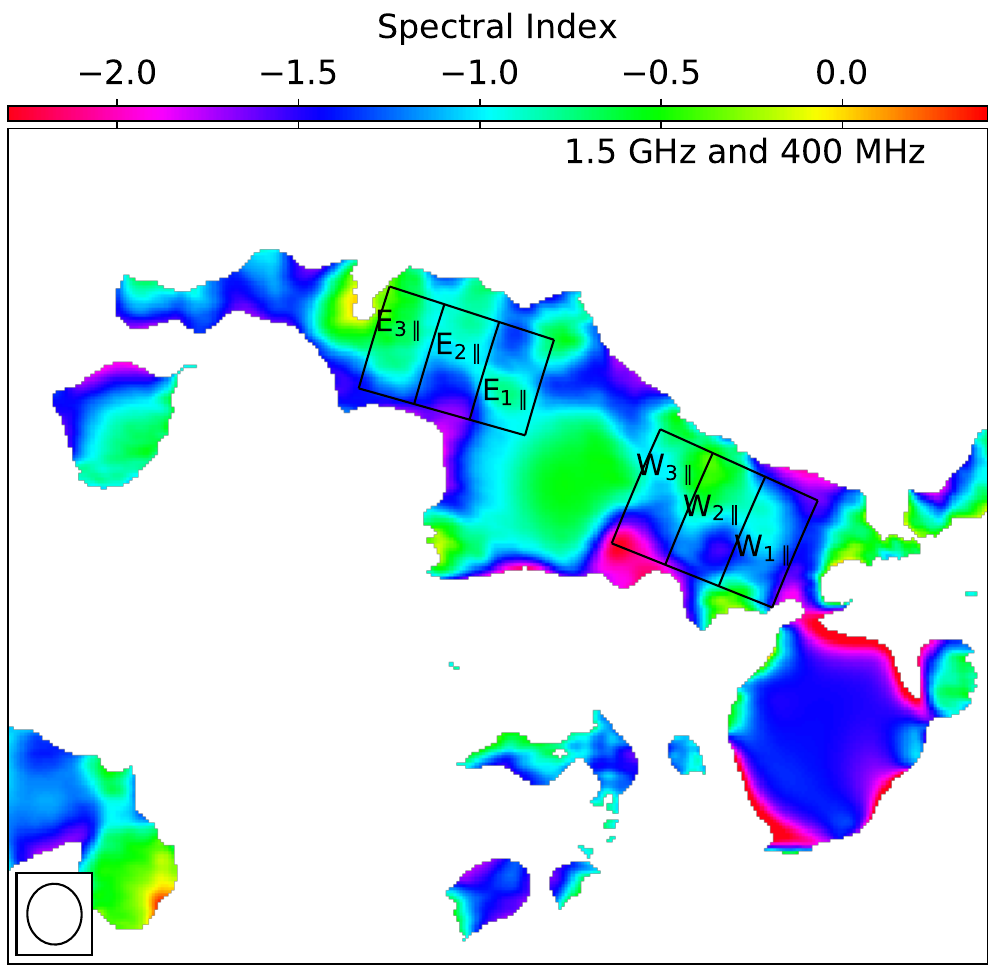}
\includegraphics[width=\columnwidth]{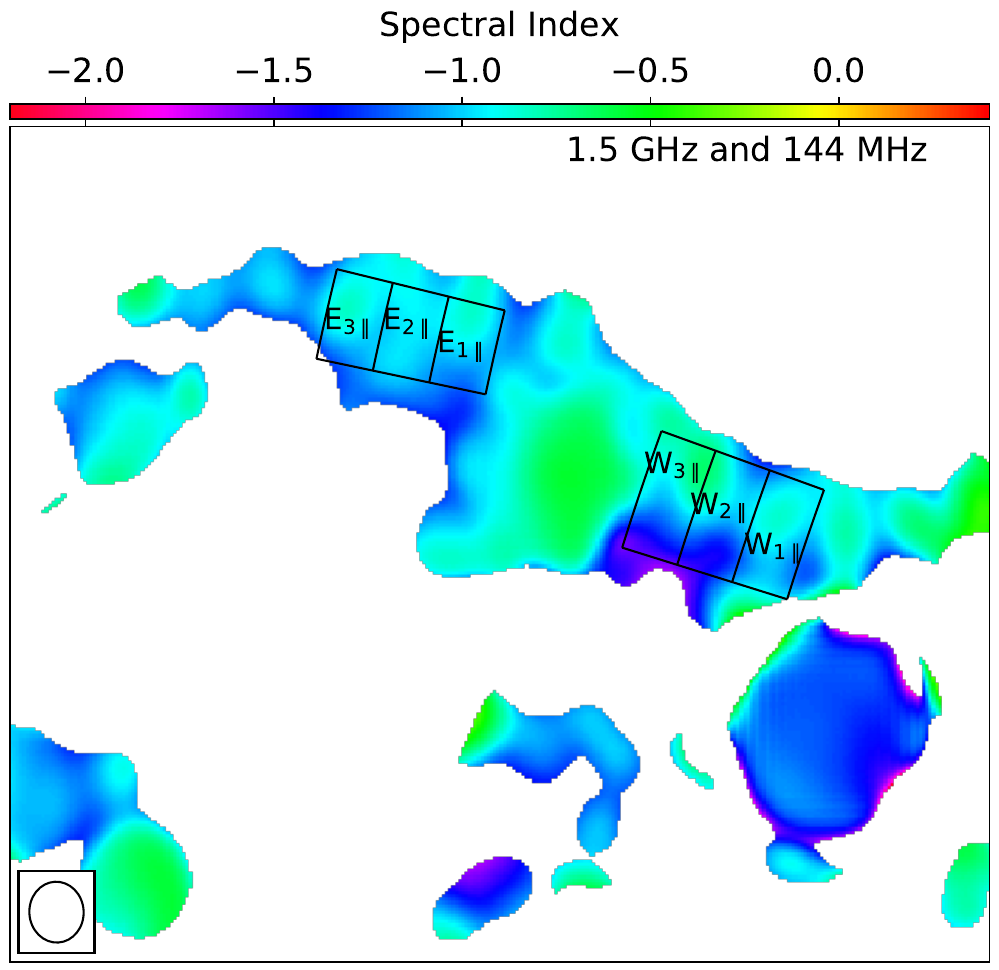}

\caption{Spectral index distribution over the R$_W$ and R$_E$ regions of the relic. Bins are placed in directions perpendicular (top panel) and parallel (bottom panel) relative to the relic's long axis.}
\label{fig: spix2}
\end{figure*}
%\subsection{Deciphering the Spectral Index Maps}
Further, we created spatially resolved spectral index maps of the radio relic to understand the particle acceleration scenario. The spectral index maps between any two frequencies and the respective error maps were created using CASA tool \textit{immath}, where we selected regions falling under 3$\sigma$ contours in both the frequencies and the remaining regions were masked. Thereafter, the spectral index values for each pixel in the selected regions were estimated using equation \ref{eq1_s} and the corresponding uncertainties were estimated using the following equation \\
\begin{equation} \label{eq2_s}
\Delta \alpha=\frac{1}{\rm{log}(\nu_1/\nu_2)}\times \sqrt{\frac{(\Delta S_1)^2}{S_1^2}+\frac{(\Delta S_2)^2}{S_2^2}}
\end{equation}

Figure~\ref{fig: spix1} illustrates the spectral index maps (a, c, e) and corresponding error maps (b, d, f) at 1.5 GHz and 400 MHz; 1.5 GHz and 144 MHz; and 400 MHz and 144 MHz, respectively.

\section{Discussion}
In the R$_W$ region of the relic, a gradual steepening of the spectral index from $\sim -0.7$ to $\sim -1.7$ is observed from the location of the detected shock by \citet{Botteon_2016MNRAS} towards the cluster centre. The spectral trend observed in A115 is characteristic of shock-generated radio relics and is consistent with observations in other relics \citep{Rajpurohit_2018ApJ, DiGennaro2018ApJ}. Here, we are particularly interested in the origin of the R$_E$ region of the relic. Notably, a spectral index gradient from the relic's outer to inner edge is also observed in the R$_E$ region, in both the $\alpha^{1.5~\mathrm{GHz}}_{400~\mathrm{MHz}}$ (Figure ~\ref{fig: spix1}a) and $\alpha^{1.5~\mathrm{GHz}}_{144~\mathrm{MHz}}$ (Figure ~\ref{fig: spix1}c) maps. Figure~\ref{fig: spix1}a shows the steepening of spectral index $\alpha^{1.5~\mathrm{GHz}}_{400~\mathrm{MHz}}$ from $\sim -0.7$ to $\sim -1.5$ as we go from outer to inner cluster region. This observation supports the idea that the R$_E$ emission also originated from the passage of the shock front. Additionally, a distinct flat spectrum trail is observed originating from the NAT galaxy. This intriguing feature suggests that electrons from the NAT galaxy may have undergone a re-energizing process triggered by the shock's passage. A similar trend of spectral index steepening at the R$_E$ region is observed in the map between 1.5 GHz and 144 MHz (Figure~\ref{fig: spix1} c). Here, the $\alpha^{1.5~\mathrm{GHz}}_{144~\mathrm{MHz}}$ varies from -0.7 to -1.4. The flat spectrum trail from the NAT galaxy is also observed here. The R$_E$ region shows a rather patchy spectral index distribution in the spectral index map between 400 MHz and 144 MHz. In the spectral index error map between 400 MHz and 144 MHz (Figure~\ref{fig: spix1}f), we observe higher uncertainty levels, particularly notable in the R$_E$ region. While the error remains minimal ($<0.1$) in the 1.5 GHz and 144 MHz map and $\leq$0.2 in the 1.5 GHz and 400 MHz map, the majority of the R$_E$ emission exhibits errors exceeding $> 0.2$ in the 400 MHz and 144 MHz map. Here, we note that LOFAR has a very dense uv-coverage at short baselines. Moreover, the lack of emission at the outer region of R$_E$ at 400 MHz can cause a steep spectral index near the outer region. We also note that the LOFAR image was not made with the same uvrange, which might induce some bias while estimating flux densities and comparing them with other frequencies. Thus, we do not use this map in our further discussion.\\
%However, there is still a hint of spectral index steepening towards the southern region of the relic.
\begin{table}
    \centering
    \caption{Spectral index distribution across bins placed in the direction perpendicular to relic's long axis for R$_W$ and R$_E$}
    \label{tab:Table2}
    \begin{subtable}{0.48\textwidth}
        \centering
        \caption{For R$_W$}
        \label{tab:Table2_1}
        \begin{tabular}{ccc} 
            \hline
            \hline
            Region & $\alpha^{1.5~\mathrm{GHz}}_{400~\mathrm{MHz}}$ & $\alpha^{1.5~\mathrm{GHz}}_{144~\mathrm{MHz}}$ \\
            \hline
            W$_{1\perp}$ & -0.89$\pm$0.09 & -0.88$\pm$0.04 \\
            W$_{2\perp}$ & -1.23$\pm$0.13 & -1.16$\pm$0.06 \\
            \hline
        \end{tabular}
    \end{subtable}
    \hfill
    \begin{subtable}{0.48\textwidth}
        \centering
        \caption{For R$_E$}
        \label{tab:Table2_2}
        \begin{tabular}{ccc} 
            \hline
            \hline
            Region & $\alpha^{1.5~\mathrm{GHz}}_{400~\mathrm{MHz}}$ & $\alpha^{1.5~\mathrm{GHz}}_{144~\mathrm{MHz}}$ \\
            \hline
            E$_{1\perp}$ & -0.89$\pm$0.12 & -0.96$\pm$0.07 \\
            E$_{2\perp}$ & -1.15$\pm$0.15 & -1.08$\pm$0.06 \\
            \hline
        \end{tabular}
    \end{subtable}
\end{table}
\subsection{Origin of Radio Relic}
There were two potential scenarios regarding the origin of the R$_E$ emission of the radio relic. As suggested by \citet{Hallman_2018ApJ}, if the R$_E$ emission originates in the presence of a shock sweeping through the radio-emitting plasma, a spectral index gradient perpendicular to the relic's long axis is expected \citep{vanWeeren_2010Sci, DiGennaro2018ApJ}. In contrast, if the radio-emitting plasma is stripped from radio galaxies and passively advected away, as suggested by \citet{Govoni_2001AA}, the spectral index gradient would align parallel to the relic's long axis and steepen opposite to the direction of the radio galaxies' motion.
\begin{table}
    \centering
    \caption{Spectral index distribution across bins placed in the direction parallel to relic's long axis for R$_W$ and R$_E$}
    \label{tab:Table3}
    \begin{subtable}{0.48\textwidth}
        \centering
        \caption{For R$_W$}
        \label{tab:Table3_1}
        \begin{tabular}{ccc} 
            \hline
            \hline
            Region & $\alpha^{1.5~\mathrm{GHz}}_{400~\mathrm{MHz}}$ & $\alpha^{1.5~\mathrm{GHz}}_{144~\mathrm{MHz}}$ \\
            \hline
            W$_{1\parallel}$ & -1.11$\pm$0.10 & -0.96$\pm$0.05 \\
            W$_{2\parallel}$ & -1.01$\pm$0.11 & -1.04$\pm$0.05 \\
            W$_{3\parallel}$ & -1.04$\pm$0.11 & -0.96$\pm$0.06 \\
            \hline
        \end{tabular}
    \end{subtable}
    \hfill
    \begin{subtable}{0.48\textwidth}
        \centering
        \caption{For R$_E$}
        \label{tab:Table3_2}
        \begin{tabular}{ccc} 
            \hline
            \hline
            Region & $\alpha^{1.5~\mathrm{GHz}}_{400~\mathrm{MHz}}$ & $\alpha^{1.5~\mathrm{GHz}}_{144~\mathrm{MHz}}$ \\
            \hline
            E$_{1\parallel}$ & -1.02$\pm$0.10 & -0.96$\pm$0.06 \\
            E$_{2\parallel}$ & -1.03$\pm$0.13 & -0.96$\pm$0.07 \\
            E$_{3\parallel}$ & -0.80$\pm$0.16 & -0.94$\pm$0.08 \\
            \hline
        \end{tabular}
    \end{subtable}
\end{table}

To understand if the entire relic follows the same spectral trend, we estimate spectral index values within bins spread in both perpendicular and parallel directions relative to the relic's long axis in the R$_W$ and R$_E$ regions (Figure~\ref{fig: spix2}). Considering the image resolution, the beam width was kept at 38$^{\prime\prime}$ and we only estimate spectral index values from the regions having a width greater than a beam size. The respective regions are marked in Figure 5, and the obtained integrated spectral indices from the regions are tabulated in Table~\ref{tab:Table2} and ~\ref{tab:Table3}. We see a significant jump in $\alpha^{1.5~\mathrm{GHz}}_{400~\mathrm{MHz}}$ and $\alpha^{1.5~\mathrm{GHz}}_{144~\mathrm{MHz}}$ perpendicular to the long axis of the relic in the R$_W$ region. This trend is anticipated, as a shock was also detected in this region, supporting the shock-induced origin of R$_W$. A noticeable jump in the R$_E$ region is also observed, suggesting a shock-induced origin of  R$_E$ (Table~\ref{tab:Table2_1}). On the other hand, across the parallel direction to the relic's long axis, no significant spectral trend was observed in the R$_W$ region (Table~\ref{tab:Table3_1}). Across the three bins, which span approximately 400 kpc at the cluster redshift in the R$_E$ region, we do not observe any significant gradual steepening away from the radio galaxy, which would be expected if the emission was caused by the radio galaxy plasma advecting away from the NAT galaxy (Table~\ref{tab:Table3_2})\citep{Krawczynski_2003MNRAS, Laing_2013MNRAS, Patra_2019Ap&SS, Velovic_2023MNRAS.523.1933V}. Rather, we observe mild flattening of spectra over 400 kpc at R$_E$ region, which can be caused by shock re-energisation of the radio galaxy plasma.
Hence, our analysis dismisses the scenario proposed by \citet{Govoni_2001AA} and supports the scenario that the R$_E$ emission was caused by shock sweeping over and re-energising the old population of radio plasma ejected from the radio galaxies near R$_E$ region \citep{Hallman_2018ApJ}.

Considering a shock-induced origin, we calculate the shock Mach number at R$_E$. The shock Mach number ($\mathcal{M}$) is related to the injection spectral index ($\alpha_{inj}$) at the location of the shock front by a non-linear equation \citep{Blandford_1987PhR}.
\begin{equation}
  \mathcal{M}_{\alpha} = \sqrt{\frac{(2\alpha_{inj}-3)}{(2\alpha_{inj}+1)}} % uses `mathrsfs`
\end{equation} 

We assess the injection index in the vicinity of the outer regions of the relic, where spectra appear relatively flatter. This area is considered a potential location for the shock front, as shocks are expected to introduce fresh particles at this position, resulting in a flat spectrum. The injection index is determined within a single beam size. With ${\alpha_{inj}}^{1.5~\mathrm{GHz}}_{400~\mathrm{MHz}}$ = -0.89 and ${\alpha_{inj}}^{1.5~\mathrm{GHz}}_{144~\mathrm{MHz}}$ = -0.96 we calculate the shock Mach number $\mathcal{M}$ ranging from 2.47$^{0.42}_{-0.24}$ to 2.31$^{0.16}_{-0.13}$. This is consistent with the shock strength at R$_W$ derived from radio observation by \citet{Hallman_2018ApJ} and further strengthens the fact that the entire relic was produced from the same shock generated by the off-axis merger. %\citet{Botteon_2016MNRAS}, with X-ray observation of the cluster, derived a Mach number $\mathcal{M}_{X-ray}$ = 1.7$\pm$0.1 (1.8$^{+0.5}_{-0.4}$) from density (temperature) jump at R$_W$ region. %The observed slight discrepancy between the radio and X-ray derived Mach numbers is anticipated. This is because radio emission depends on the magnitude of magnetic field fluctuations, which decline at a slower rate compared to density and temperature fluctuations. Consequently, radio-derived Mach numbers tend to skew towards the higher end of the Mach number spectrum \citep{Datta_2014ApJ...793...80D, Wittor_2021MNRAS.506..396W}. Furthermore, the higher difference could suggest that additional processes, such as the re-acceleration of electrons from radio galaxies, play a role in the generation of the relic \citep{Colafrancesco_2017MNRAS}. 
In the case of A115, it is notable that the shock is too weak to accelerate thermal particles to relativistic speed at R$_E$, where the X-ray luminosity is low. Moreover, the lack of X-ray emission near the R$_E$ region rules out the possibility of the origin of R$_E$ from only the shock acceleration of thermal particles \citep{Botteon_2016MNRAS}. This strengthens the idea that the shock re-accelerated the aged population of radio galaxy plasma at R$_E$ region.

\section{Conclusion}
We performed a multi-frequency study on the peculiar radio relic in A115 using LOFAR 144 MHz, uGMRT 400 MHz and VLA 1.5 GHz observations. The gradual steepening of the relic's spectral index from the cluster outskirts seen in our study aligns with patterns seen in other instances of shock-generated radio relics \citep{Rajpurohit_2018ApJ, DiGennaro2018ApJ}. The similar spectral trend and the comparable strength of the shock observed at both the eastern and western regions of the relic suggest that the entire relic likely originated from a single shock event. Furthermore, the absence of thermal emission near the eastern region and the observed spectral index trend indicate the significant contribution of re-energized fossil electrons from radio galaxies in shaping the relic's structure. Our analysis supports the notion that, while the primary driver of relic emission is the shock resulting from an off-axis merger, as proposed by \citet{Hallman_2018ApJ} and \citet{Lee_2020ApJ}, the presence of fossil electrons from radio galaxies near the relic's eastern region plays a crucial role in defining its overall morphology.

\section{Acknowledgements}

We thank the anonymous reviewer for their comments and suggestions. We thank IIT Indore for providing us with the necessary means to conduct the research. We acknowledge the use of facilities procured through the funding via the Department of Science and Technology, Government of India sponsored DST-FIST grant no. SR/FST/PSII/2021/162 (C) awarded to the DAASE, IIT Indore. We thank the staff of the GMRT who have made these observations possible. The GMRT is run by the National Centre for Radio Astrophysics of the Tata Institute of Fundamental Research. This research made use of Astropy,\footnote{http://www.astropy.org} a community-developed core Python package for Astronomy \citep{Astropy_2013AA, astropy_2018}, Matplotlib \citep{Matplotlib_Hunter:2007}, and APLpy, an open-source plotting package for Python \citep{Robitaille_APLpy_2012}.

\bibliography{example}

\begin{thebibliography}{}
\expandafter\ifx\csname natexlab\endcsname\relax\def\natexlab#1{#1}\fi

\bibitem[{and A.~M. Price-Whelan {$et~al$.}(2018)and A.~M. Price-Whelan, Sip{\H{o}}cz, Günther, Lim, Crawford, Conseil, Shupe, Craig, Dencheva, Ginsburg, VanderPlas, Bradley, P{\'{e}}rez-Su{\'{a}}rez, de~Val-Borro, Aldcroft, Cruz, Robitaille, Tollerud, Ardelean, Babej, Bach, Bachetti, Bakanov, Bamford, Barentsen, Barmby, Baumbach, Berry, Biscani, Boquien, Bostroem, Bouma, Brammer, Bray, Breytenbach, Buddelmeijer, Burke, Calderone, Rodr{\'{\i}}guez, Cara, Cardoso, Cheedella, Copin, Corrales, Crichton, D'Avella, Deil, Depagne, Dietrich, Donath, Droettboom, Earl, Erben, Fabbro, Ferreira, Finethy, Fox, Garrison, Gibbons, Goldstein, Gommers, Greco, Greenfield, Groener, Grollier, Hagen, Hirst, Homeier, Horton, Hosseinzadeh, Hu, Hunkeler, Ivezi{\'{c}}, Jain, Jenness, Kanarek, Kendrew, Kern, Kerzendorf, Khvalko, King, Kirkby, Kulkarni, Kumar, Lee, Lenz, Littlefair, Ma, Macleod, Mastropietro, McCully, Montagnac, Morris, Mueller, Mumford, Muna, Murphy, Nelson, Nguyen, Ninan, Nöthe, Ogaz, Oh, Parejko, Parley, Pascual,
  Patil, Patil, Plunkett, Prochaska, Rastogi, Janga, Sabater, Sakurikar, Seifert, Sherbert, Sherwood-Taylor, Shih, Sick, Silbiger, Singanamalla, Singer, Sladen, Sooley, Sornarajah, Streicher, Teuben, Thomas, Tremblay, Turner, Terr{\'{o}}n, van Kerkwijk, de~la Vega, Watkins, Weaver, Whitmore, Woillez, Zabalza, , \& and}]{astropy_2018}
and A.~M. Price-Whelan, Sip{\H{o}}cz, B.~M., Günther, H.~M., {$et~al$.} 2018, The Astronomical Journal, 156, 123

\bibitem[{{Astropy Collaboration} {$et~al$.}(2013){Astropy Collaboration}, {Robitaille}, {Tollerud}, {Greenfield}, {Droettboom}, {Bray}, {Aldcroft}, {Davis}, {Ginsburg}, {Price-Whelan}, {Kerzendorf}, {Conley}, {Crighton}, {Barbary}, {Muna}, {Ferguson}, {Grollier}, {Parikh}, {Nair}, {Unther}, {Deil}, {Woillez}, {Conseil}, {Kramer}, {Turner}, {Singer}, {Fox}, {Weaver}, {Zabalza}, {Edwards}, {Azalee Bostroem}, {Burke}, {Casey}, {Crawford}, {Dencheva}, {Ely}, {Jenness}, {Labrie}, {Lim}, {Pierfederici}, {Pontzen}, {Ptak}, {Refsdal}, {Servillat}, \& {Streicher}}]{Astropy_2013AA}
{Astropy Collaboration}, {Robitaille}, T.~P., {Tollerud}, E.~J., {$et~al$.} 2013, aap, 558, A33

\bibitem[{{Barrena} {$et~al$.}(2007){Barrena}, {Boschin}, {Girardi}, \& {Spolaor}}]{Barrena_2007AA}
{Barrena}, R., {Boschin}, W., {Girardi}, M., \& {Spolaor}, M. 2007, aap, 469, 861

\bibitem[{{Blandford} \& {Eichler}(1987)}]{Blandford_1987PhR}
{Blandford}, R., \& {Eichler}, D. 1987, physrep, 154, 1

\bibitem[{{Botteon} {$et~al$.}(2020){Botteon}, {Brunetti}, {Ryu}, \& {Roh}}]{Botteon_2020AA}
{Botteon}, A., {Brunetti}, G., {Ryu}, D., \& {Roh}, S. 2020, aap, 634, A64

\bibitem[{{Botteon} {$et~al$.}(2016{\natexlab{a}}){Botteon}, {Gastaldello}, {Brunetti}, \& {Dallacasa}}]{Botteon_2016MNRAS}
{Botteon}, A., {Gastaldello}, F., {Brunetti}, G., \& {Dallacasa}, D. 2016{\natexlab{a}}, mnras, 460, L84

\bibitem[{{Botteon} {$et~al$.}(2016{\natexlab{b}}){Botteon}, {Gastaldello}, {Brunetti}, \& {Kale}}]{Botteon_2016MNRASB}
{Botteon}, A., {Gastaldello}, F., {Brunetti}, G., \& {Kale}, R. 2016{\natexlab{b}}, mnras, 463, 1534

\bibitem[{{Botteon} {$et~al$.}(2022){Botteon}, {Shimwell}, {Cassano}, {Cuciti}, {Zhang}, {Bruno}, {Camillini}, {Natale}, {Jones}, {Gastaldello}, {Simionescu}, {Rossetti}, {Akamatsu}, {van Weeren}, {Brunetti}, {Br{\"u}ggen}, {Groeneveld}, {Hoang}, {Hardcastle}, {Ignesti}, {Di Gennaro}, {Bonafede}, {Drabent}, {R{\"o}ttgering}, {Hoeft}, \& {de Gasperin}}]{Botteon_2022AA}
{Botteon}, A., {Shimwell}, T.~W., {Cassano}, R., {$et~al$.} 2022, aap, 660, A78

\bibitem[{{Bourdin} {$et~al$.}(2013){Bourdin}, {Mazzotta}, {Markevitch}, {Giacintucci}, \& {Brunetti}}]{Bourdin_2013ApJ}
{Bourdin}, H., {Mazzotta}, P., {Markevitch}, M., {Giacintucci}, S., \& {Brunetti}, G. 2013, apj, 764, 82

\bibitem[{{Brunetti} \& {Jones}(2014)}]{Brunetti_2014IJMPD..2330007B}
{Brunetti}, G., \& {Jones}, T.~W. 2014, International Journal of Modern Physics D, 23, 1430007

\bibitem[{{Chatterjee} {$et~al$.}(2024){Chatterjee}, {Rahaman}, {Datta}, {Kale}, \& {Paul}}]{Chatterjee_2024MNRAS}
{Chatterjee}, S., {Rahaman}, M., {Datta}, A., {Kale}, R., \& {Paul}, S. 2024, mnras, 527, 10986

\bibitem[{{Chatterjee} {$et~al$.}(2022){Chatterjee}, {Rahaman}, {Datta}, \& {Raja}}]{chatterjee_2022AJ}
{Chatterjee}, S., {Rahaman}, M., {Datta}, A., \& {Raja}, R. 2022, aj, 164, 83

\bibitem[{{Di Gennaro} {$et~al$.}(2018){Di Gennaro}, {van Weeren}, {Hoeft}, {Kang}, {Ryu}, {Rudnick}, {Forman}, {R{\"o}ttgering}, {Br{\"u}ggen}, {Dawson}, {Golovich}, {Hoang}, {Intema}, {Jones}, {Kraft}, {Shimwell}, \& {Stroe}}]{DiGennaro2018ApJ}
{Di Gennaro}, G., {van Weeren}, R.~J., {Hoeft}, M., {$et~al$.} 2018, apj, 865, 24

\bibitem[{{Ensslin} {$et~al$.}(1998){Ensslin}, {Biermann}, {Klein}, \& {Kohle}}]{Ensslin_1998A&A}
{Ensslin}, T.~A., {Biermann}, P.~L., {Klein}, U., \& {Kohle}, S. 1998, aap, 332, 395

\bibitem[{{Forman} {$et~al$.}(1981){Forman}, {Bechtold}, {Blair}, {Giacconi}, {van Speybroeck}, \& {Jones}}]{Forman_1981ApJ}
{Forman}, W., {Bechtold}, J., {Blair}, W., {$et~al$.} 1981, apjl, 243, L133

\bibitem[{Gennaro {$et~al$.}(2018)Gennaro, van Weeren, Hoeft, Kang, Ryu, Rudnick, Forman, Röttgering, Brüggen, Dawson, Golovich, Hoang, Intema, Jones, Kraft, Shimwell, \& Stroe}]{Gennaro_2018}
Gennaro, G.~D., van Weeren, R.~J., Hoeft, M., {$et~al$.} 2018, The Astrophysical Journal, 865, 24

\bibitem[{{Govoni} {$et~al$.}(2001){Govoni}, {Feretti}, {Giovannini}, {B{\"o}hringer}, {Reiprich}, \& {Murgia}}]{Govoni_2001AA}
{Govoni}, F., {Feretti}, L., {Giovannini}, G., {$et~al$.} 2001, aap, 376, 803

\bibitem[{{Hallman} {$et~al$.}(2018){Hallman}, {Alden}, {Rapetti}, {Datta}, \& {Burns}}]{Hallman_2018ApJ}
{Hallman}, E.~J., {Alden}, B., {Rapetti}, D., {Datta}, A., \& {Burns}, J.~O. 2018, apj, 859, 44

\bibitem[{Hunter(2007)}]{Matplotlib_Hunter:2007}
Hunter, J.~D. 2007, Computing in Science \& Engineering, 9, 90

\bibitem[{{Intema} {$et~al$.}(2017){Intema}, {Jagannathan, P.}, {Mooley, K. P.}, \& {Frail, D. A.}}]{Intema_refId0}
{Intema}, {Jagannathan, P.}, {Mooley, K. P.}, \& {Frail, D. A.} 2017, A\&A, 598, A78

\bibitem[{{Kim} {$et~al$.}(2019){Kim}, {Jee}, {Finner}, {Golovich}, {Wittman}, {van Weeren}, \& {Dawson}}]{Kim_2019ApJ}
{Kim}, M., {Jee}, M.~J., {Finner}, K., {$et~al$.} 2019, apj, 874, 143

\bibitem[{{Krawczynski} {$et~al$.}(2003){Krawczynski}, {Harris}, {Grossman}, {Lane}, {Kassim}, \& {Willis}}]{Krawczynski_2003MNRAS}
{Krawczynski}, H., {Harris}, D.~E., {Grossman}, R., {$et~al$.} 2003, mnras, 345, 1255

\bibitem[{{Laing} \& {Bridle}(2013)}]{Laing_2013MNRAS}
{Laing}, R.~A., \& {Bridle}, A.~H. 2013, mnras, 432, 1114

\bibitem[{{Lee} {$et~al$.}(2020){Lee}, {Jee}, {Kang}, {Ryu}, {Kimm}, \& {Br{\"u}ggen}}]{Lee_2020ApJ}
{Lee}, W., {Jee}, M.~J., {Kang}, H., {$et~al$.} 2020, apj, 894, 60

\bibitem[{{Offringa} \& {Smirnov}(2017)}]{Offringa_2017MNRAS}
{Offringa}, A.~R., \& {Smirnov}, O. 2017, mnras, 471, 301

\bibitem[{{Patra} {$et~al$.}(2019){Patra}, {Pal}, {Konar}, \& {Chakrabarti}}]{Patra_2019Ap&SS}
{Patra}, D., {Pal}, S., {Konar}, C., \& {Chakrabarti}, S.~K. 2019, apss, 364, 72

\bibitem[{{Paul} {$et~al$.}(2023){Paul}, {Kale}, {Datta}, {Basu}, {Sur}, {Parekh}, {Gupta}, {Chatterjee}, {Salunkhe}, {Iqbal}, {Pandey-Pommier}, {Raja}, {Rahaman}, {Raychaudhury}, {Nath}, \& {Majumdar}}]{paul_2023JApA}
{Paul}, S., {Kale}, R., {Datta}, A., {$et~al$.} 2023, Journal of Astrophysics and Astronomy, 44, 38

\bibitem[{{Planck Collaboration} {$et~al$.}(2014){Planck Collaboration}, {Ade}, {Aghanim}, {Armitage-Caplan}, {Arnaud}, {Ashdown}, {Atrio-Barandela}, {Aumont}, {Aussel}, {Baccigalupi}, {Banday}, {Barreiro}, {Barrena}, {Bartelmann}, {Bartlett}, {Battaner}, {Benabed}, {Beno{\^\i}t}, {Benoit-L{\'e}vy}, {Bernard}, {Bersanelli}, {Bielewicz}, {Bikmaev}, {Bobin}, {Bock}, {B{\"o}hringer}, {Bonaldi}, {Bond}, {Borrill}, {Bouchet}, {Bridges}, {Bucher}, {Burenin}, {Burigana}, {Butler}, {Cardoso}, {Carvalho}, {Catalano}, {Challinor}, {Chamballu}, {Chary}, {Chen}, {Chiang}, {Chiang}, {Chon}, {Christensen}, {Churazov}, {Church}, {Clements}, {Colombi}, {Colombo}, {Comis}, {Couchot}, {Coulais}, {Crill}, {Curto}, {Cuttaia}, {Da Silva}, {Dahle}, {Danese}, {Davies}, {Davis}, {de Bernardis}, {de Rosa}, {de Zotti}, {Delabrouille}, {Delouis}, {D{\'e}mocl{\`e}s}, {D{\'e}sert}, {Dickinson}, {Diego}, {Dolag}, {Dole}, {Donzelli}, {Dor{\'e}}, {Douspis}, {Dupac}, {Efstathiou}, {Eisenhardt}, {En{\ss}lin}, {Eriksen}, {Feroz}, {Finelli},
  {Flores-Cacho}, {Forni}, {Frailis}, {Franceschi}, {Fromenteau}, {Galeotta}, {Ganga}, {G{\'e}nova-Santos}, {Giard}, {Giardino}, {Gilfanov}, {Giraud-H{\'e}raud}, {Gonz{\'a}lez-Nuevo}, {G{\'o}rski}, {Grainge}, {Gratton}, {Gregorio}, {Groeneboom}, {Gruppuso}, {Hansen}, {Hanson}, {Harrison}, {Hempel}, {Henrot-Versill{\'e}}, {Hern{\'a}ndez-Monteagudo}, {Herranz}, {Hildebrandt}, {Hivon}, {Hobson}, {Holmes}, {Hornstrup}, {Hovest}, {Huffenberger}, {Hurier}, {Hurley-Walker}, {Jaffe}, {Jaffe}, {Jones}, {Juvela}, {Keih{\"a}nen}, {Keskitalo}, {Khamitov}, {Kisner}, {Kneissl}, {Knoche}, {Knox}, {Kunz}, {Kurki-Suonio}, {Lagache}, {L{\"a}hteenm{\"a}ki}, {Lamarre}, {Lasenby}, {Laureijs}, {Lawrence}, {Leahy}, {Leonardi}, {Le{\'o}n-Tavares}, {Lesgourgues}, {Li}, {Liddle}, {Liguori}, {Lilje}, {Linden-V{\o}rnle}, {L{\'o}pez-Caniego}, {Lubin}, {Mac{\'\i}as-P{\'e}rez}, {MacTavish}, {Maffei}, {Maino}, {Mandolesi}, {Maris}, {Marshall}, {Martin}, {Mart{\'\i}nez-Gonz{\'a}lez}, {Masi}, {Massardi}, {Matarrese}, {Matthai}, {Mazzotta},
  {Mei}, {Meinhold}, {Melchiorri}, {Melin}, {Mendes}, {Mennella}, {Migliaccio}, {Mikkelsen}, {Mitra}, {Miville-Desch{\^e}nes}, {Moneti}, {Montier}, {Morgante}, {Mortlock}, {Munshi}, {Murphy}, {Naselsky}, {Nati}, {Natoli}, {Nesvadba}, {Netterfield}, {N{\o}rgaard-Nielsen}, {Noviello}, {Novikov}, {Novikov}, {O'Dwyer}, {Olamaie}, {Osborne}, {Oxborrow}, {Paci}, {Pagano}, {Pajot}, {Paoletti}, {Pasian}, {Patanchon}, {Pearson}, {Perdereau}, {Perotto}, {Perrott}, {Perrotta}, {Piacentini}, {Piat}, {Pierpaoli}, {Pietrobon}, {Plaszczynski}, {Pointecouteau}, {Polenta}, {Ponthieu}, {Popa}, {Poutanen}, {Pratt}, {Pr{\'e}zeau}, {Prunet}, {Puget}, {Rachen}, {Reach}, {Rebolo}, {Reinecke}, {Remazeilles}, {Renault}, {Ricciardi}, {Riller}, {Ristorcelli}, {Rocha}, {Rosset}, {Roudier}, {Rowan-Robinson}, {Rubi{\~n}o-Mart{\'\i}n}, {Rumsey}, {Rusholme}, {Sandri}, {Santos}, {Saunders}, {Savini}, {Schammel}, {Scott}, {Seiffert}, {Shellard}, {Shimwell}, {Spencer}, {Stanford}, {Starck}, {Stolyarov}, {Stompor}, {Sudiwala}, {Sunyaev},
  {Sureau}, {Sutton}, {Suur-Uski}, {Sygnet}, {Tauber}, {Tavagnacco}, {Terenzi}, {Toffolatti}, {Tomasi}, {Tristram}, {Tucci}, {Tuovinen}, {T{\"u}rler}, {Umana}, {Valenziano}, {Valiviita}, {Van Tent}, {Vibert}, {Vielva}, {Villa}, {Vittorio}, {Wade}, {Wandelt}, {White}, {White}, {Yvon}, {Zacchei}, \& {Zonca}}]{Planck_Collaboration_2014AA}
{Planck Collaboration}, {Ade}, P.~A.~R., {Aghanim}, N., {$et~al$.} 2014, aap, 571, A29

\bibitem[{{Raja} {$et~al$.}(2023){Raja}, {Rahaman}, {Datta}, \& {Smirnov}}]{Raja_2023arXiv}
{Raja}, R., {Rahaman}, M., {Datta}, A., \& {Smirnov}, O.~M. 2023, arXiv e-prints, arXiv:2309.14244

\bibitem[{{Rajpurohit} {$et~al$.}(2018){Rajpurohit}, {Hoeft}, {van Weeren}, {Rudnick}, {R{\"o}ttgering}, {Forman}, {Br{\"u}ggen}, {Croston}, {Andrade-Santos}, {Dawson}, {Intema}, {Kraft}, {Jones}, \& {Jee}}]{Rajpurohit_2018ApJ}
{Rajpurohit}, K., {Hoeft}, M., {van Weeren}, R.~J., {$et~al$.} 2018, apj, 852, 65

\bibitem[{{Robitaille} \& {Bressert}(2012)}]{Robitaille_APLpy_2012}
{Robitaille}, T., \& {Bressert}, E. 2012, {APLpy: Astronomical Plotting Library in Python}, ascl:1208.017

\bibitem[{Sarazin(2002)}]{Sarazin_2002}
Sarazin, C.~L. 2002, Merging Processes in Galaxy Clusters, 1–38

\bibitem[{Scaife \& Heald(2012)}]{scaife_10.1111/j.1745-3933.2012.01251.x}
Scaife, A. M.~M., \& Heald, G.~H. 2012, Monthly Notices of the Royal Astronomical Society: Letters, 423, L30

\bibitem[{{van Weeren} {$et~al$.}(2019){van Weeren}, {de Gasperin}, {Akamatsu}, {Br{\"u}ggen}, {Feretti}, {Kang}, {Stroe}, \& {Zandanel}}]{vanweeren_2019SSRv..215...16V}
{van Weeren}, R.~J., {de Gasperin}, F., {Akamatsu}, H., {$et~al$.} 2019, ssr, 215, 16

\bibitem[{{van Weeren} {$et~al$.}(2010){van Weeren}, {R{\"o}ttgering}, {Br{\"u}ggen}, \& {Hoeft}}]{vanWeeren_2010Sci}
{van Weeren}, R.~J., {R{\"o}ttgering}, H. J.~A., {Br{\"u}ggen}, M., \& {Hoeft}, M. 2010, Science, 330, 347

\bibitem[{{Velovi{\'c}} {$et~al$.}(2023){Velovi{\'c}}, {Cotton}, {Filipovi{\'c}}, {Norris}, {Barnes}, \& {Condon}}]{Velovic_2023MNRAS.523.1933V}
{Velovi{\'c}}, V., {Cotton}, W.~D., {Filipovi{\'c}}, M.~D., {$et~al$.} 2023, mnras, 523, 1933

\end{thebibliography}

%%Use section* for acknowledgements

%%use \balance somewhere in the left column of the last page to balance the two columns in the end page

%\bibitem{latexcompanion}
%Clark D. H., Caswell J. L. 1976, MNRAS, 174, 267

\end{document}